\documentclass[useAMS,usenatbib]{mn2e}
\usepackage{times}
\usepackage{amssymb}
\usepackage{graphicx}
\usepackage{rotating}
\usepackage{lscape}
\usepackage{multirow}
\usepackage{float}
\usepackage{fixltx2e}

\begin{document}

%
%
%
%



\title[BCG AGN feedback and the ICM]{The XMM Cluster Survey: The interplay between the brightest cluster galaxy and the intra-cluster medium via AGN feedback.} 
\author[J. P. Stott et al.]{John P. Stott$^{1}$\thanks{E-mail:
j.p.stott@durham.ac.uk}, Ryan C. Hickox$^{1}$, Alastair C. Edge$^{1}$, Chris A. Collins$^{2}$, Matt Hilton$^{3}$,  \newauthor Craig D. Harrison$^{4}$, A. Kathy Romer$^{5}$, Philip J. Rooney$^{5}$, Scott T. Kay$^{6}$, \newauthor Christopher J. Miller$^{4}$, Martin Sahl{\'e}n$^{7}$,  Ed J. Lloyd-Davies$^{5}$, Nicola Mehrtens$^{5}$, \newauthor Ben Hoyle$^{8,9,10}$, Andrew R. Liddle$^{5}$, Pedro T. P. Viana$^{11,12}$, Ian G. McCarthy$^{13,14}$, \newauthor Joop Schaye$^{15}$, C. M. Booth$^{15}$
\\
$^{1}$Extragalactic \& Cosmology Group, Department of Physics, University of Durham, South Road, Durham DH1 3LE, UK\\
$^{2}$Astrophysics Research Institute, Liverpool John Moores University, Twelve Quays House, Egerton Wharf, Birkenhead CH41 1LD, UK\\
$^3$ School of Physics \& Astronomy, University of Nottingham, Nottingham, NG7 2RD, UK\\
$^4$ Department of Astronomy, University of Michigan, Ann Arbor, MI 48109, USA\\
$^5$ Astronomy Centre, University of Sussex, Falmer, Brighton, BN1 9QH, UK\\
$^6$ Jodrell Bank Centre for Astrophysics, School of Physics and Astronomy, The University of Manchester, Manchester M13 9PL\\
$^{7}$ The Oskar Klein Centre for Cosmoparticle Physics, Department of Physics, Stockholm University, AlbaNova, SE-106 91, Stockholm, Sweden\\
$^8$ Institute of Sciences of the Cosmos (ICCUB) and IEEC, Physics Department, University of Barcelona, Barcelona 08024,
  Spain\\
$^9$ CSIC, Consejo Superior de Investigaciones Cientificas, Serrano 117, Madrid, 28006, Spain\\
$^{10}$ Helsinki Institute of Physics, P.O. Box 64, FIN-00014 University of Helsinki, Finland\\
$^{11}$ Centro de Astrof\'{\i}sica da Universidade do Porto, Rua das Estrelas, 4150-762, Porto, Portugal\\
$^{12}$ Departamento de F\'{\i}sica e Astronomia da Faculdade de Ci\^{e}ncias da Universidade do Porto, Rua do Campo Alegre, 687, 4169-007 Porto, Portugal\\
$^{13}$Kavli Institute for Cosmology, University of Cambridge, Madingley Road, Cambridge CB3 0HA, UK\\
$^{14}$Astrophysics and Space Research Group, School of Physics \& Astronomy, University of Birmingham, Edgbaston, Birmingham, B15 2TT, UK\\
$^{15}$Leiden Observatory, Leiden University, P.O. Box 9513, 2300 RA Leiden, The Netherlands}

\date{}

\pagerange{\pageref{firstpage}--\pageref{lastpage}} \pubyear{2002}

\maketitle

\label{firstpage}

\begin{abstract}

Using a sample of 123 X-ray clusters and groups drawn from the XMM-Cluster Survey first data release, we investigate the interplay between the brightest cluster galaxy (BCG), its black hole, and the intra-cluster/group medium (ICM). It appears that for groups and clusters with a BCG likely to host significant AGN feedback, gas cooling dominates in those with $T_{X}>2\,\rm keV$ while AGN feedback dominates below. This may be understood through the sub-unity exponent found in the scaling relation we derive between the BCG mass and cluster mass over the halo mass range $10^{13}<M_{500}<10^{15}\rm M_{\odot}$ and the lack of correlation between radio luminosity and cluster mass, such that BCG AGN in groups can have relatively more energetic influence on the ICM. The $L_{X}-T_{X}$ relation for systems with the most massive BCGs, or those with BCGs co-located with the peak of the ICM emission, is steeper than that for those with the least massive and most offset, which instead follows self-similarity. This is evidence that a combination of central gas cooling and powerful, well fuelled AGN causes the departure of the ICM from pure gravitational heating, with the steepened relation crossing self-similarity at $T_{X}=2\,\rm keV$. Importantly, regardless of their black hole mass, BCGs are more likely to host radio-loud AGN if they are in a massive cluster ($T_{X}\gtrsim2\,\rm keV$) and again co-located with an effective fuel supply of dense, cooling gas. This demonstrates that the most massive black holes appear to know more about their host cluster than they do about their host galaxy. The results lead us to propose a physically motivated, empirical definition of `cluster' and `group', delineated at 2\,keV. 

\end{abstract}


\section{Introduction}

Galaxy clusters and groups are important probes of cosmology as they trace the densest regions of the Universe, which are the descendants of the first over-densities to collapse after the big bang. They can be parameterised by the X-ray emission properties of their hot intra-cluster medium (ICM), with both the luminosity of the X-ray emission ($L_{X}$) and the temperature of the gas ($T_{X}$) scaling with the total mass of the system (e.g. \citealt{pratt2009}). $L_{X}$ also scales with the integral of the square of the gas density while $T_{X}$ indicates the virial temperature of the system. For a self-similar relation, where gravitational heating is the only source of energy, $L_{X}\propto T_{X}^{2}$ \citep{kaiser1986} although the observed slope is typically $2-3$ \citep{mushotzky1984,edge1991b,markevitch1998,arnaud1999,pratt2009,mittal2011} which demonstrates that secondary and non-gravitational effects are influencing the ICM \citep{allen1998,muanwong2002,randall2002,rowley2004,mag2007,hartley2008}. 

As both $L_{X}$ and $T_{X}$ are used to calculate cluster mass but probe different physical properties of the gas, it is possible to assess the state of the ICM by studying the relation between them for a sample of clusters. It is important to understand this $L_{X} - T_{X}$ relation and any definable sub-samples within it, as X-ray mass scaling relations are used to derive cluster masses and are employed as complementary probes of cosmology to those derived from other methods \citep{borgani2001,stanek2006,allen2008,vik2009,sahlen2009}. A further motivation is to explore the physical processes that act within the ICM itself. For example, we seek to understand the process that provides an entropy excess to stop the gas in the centres of clusters from catastrophically over-cooling via the classical cooling flow model \citep{fabian1994}. The leading candidate for this is the energy output by active galactic nuclei (AGN, \citealt{tucker1997,mcnamara2007,gitti2011}). This is also often invoked in one form or another to reconcile the dark matter mass function with the galaxy luminosity function \citep{best2006,croton2006,bower2006} and is thus a cosmologically important energy source across a range of mass scales. 

In terms of gas cooling, for dynamically relaxed clusters that have a cool-core present (i.e. the cooling time is significantly less than a Hubble time) the central density of gas and thus $L_{X}$ will be relatively high  \citep{fabian1994}. This manifests itself in part through an offset or steepening of the $L_{X}-T_{X}$ relation for cool-core clusters relative to the whole population \citep{fabian1994b,mittal2011}. However when corrected using both X-ray spectra and imaging or ignoring the central regions, the self-similar result can be recovered for some clusters \citep{allen1998,maughan2011}. In contrast, unvirialised systems, for example those undergoing a cluster-cluster merger, will have disturbed X-ray emission and a $T_{X}$ that is less indicative of the total mass of the system \citep{randall2002,rowley2004}. 

As mentioned above, the galaxies themselves can influence the gas through non-gravitational processes due to the presence of AGN which can heat and even blow the gas out of the central regions of clusters and groups, and their progenitors. The effect of radio-loud AGN activity on the ICM has been seen in a number of detailed X-ray and radio observations of local clusters and groups where the radio lobes of the AGN are found to inflate large cavities in the X-ray emitting gas, demonstrating the power of this process \citep{churazov2001,birzan2004}. In the cluster mass regime the power in these AGN cavities matches the luminosity of the central cooling ICM, whereas in low mass clusters and groups the cavity power is found to exceed that of the ICM, and thus the energy injection from AGN can dominate in the core \citep{gitti2011}. AGN processes such as these can lower the central gas density of clusters, and therefore $L_{X}$, while the $T_{X}$ is less affected, as is the case in simulations \citep{puchwein2008,mccarthy2010}. Alternatively this can be thought of as heating the gas for a given $L_{X}$ as has been reported observationally \citep{croston2005}. 

There is further observational evidence connecting the galaxy scale to the cluster scale, as the broad-band optical and near-infrared properties of the central galaxies in groups and clusters (hereafter BCGs) have been found to correlate with the X-ray luminosity ($ L_{X}$) and X-ray temperature ($ T_{X}$) of the hot ICM of their hosts, albeit with significant scatter \citep{edge1991,collins1998,lin2004,popesso2007,whiley2008,stott2008,brough2008,mittal2009}. As these X-ray properties correlate to first order with the total mass of the system, BCG stellar mass must correlate with the dark matter halo mass of the cluster. This may be indicative of the hierarchical build up of the galaxies and their host halos over cosmic time, as these masses are also found to correlate in theoretical models \citep{delucia2007}. 

We may therefore ask, are the BCGs influenced by and influencing the ICM, and if so how? We can infer from the $M_{\rm BH} - \sigma$ relation, in which black hole mass is found to correlate with the host galaxy velocity dispersion \citep{ferrarese2000,gebhardt2000}, and the relationship between black hole mass and galaxy bulge mass ($M_{\rm BH} - M_{\rm Bulge}$, \citealt{magorrian1998}), that all BCGs contain supermassive black holes. However because of the duty-cycle of supermassive black hole accretion, only a certain fraction are AGN at any one instant. Observationally, cluster galaxies and BCGs in particular are more likely to host AGN and these are more likely to be powerful radio sources, having the highest duty-cycles of any class of galaxy, with duty-cycle correlating with stellar mass \citep{lin2007,best2007}. Clustering and environment studies of radio-loud AGN
consistently find that they typically reside in halos of mass $> 10^{13} \rm M_{\odot}$ \citep[e.g.,][]{best04radio, wake08radio, mand09agnclust, hick09corr, dono10clust}, precisely the halos with virialised hot atmospheres for which AGN heating is required in cosmological models. There is also a correlation between BCG stellar mass and the maximum radio luminosity observed for that mass of galaxy, albeit with large scatter \citep{lin2007}. This can be understood in terms of the $M_{\rm BH} - M_{\rm Bulge}$ relation, and the fact that AGN power output correlates with black hole accretion rate, which is a function of black hole mass (e.g. Eddington or Bondi accretion, \citealt{bondi1952}). The significant scatter is due to observing the AGN at different stages of the accretion cycle between observationally quiescent and radio-loud AGN. One might therefore expect that the most massive BCGs will have supermassive black holes able to inject the most energy into their surroundings and thus most strongly influence the ICM.  

This may imply that clusters and groups containing the most massive central galaxies will have a lower $L_{X}$ for a given $T_{X}$, as they will be those that possess black holes able to inject the most energy into the ICM. However, the process of AGN accretion and then energy injection into its surroundings is often described as a feedback process that regulates its own accretion rate by heating and/or mechanically removing its own fuelling gas supply \citep{best2006,bower2006,mcnamara2007,booth2010}. In-falling gas fuels the accreting black hole, producing radiation driven winds from the accretion disk or mechanical jets of radio emitting material, that act to stop the gas from the surrounding medium from cooling onto the accretion disk. This requires the energy output to be coupled in some way to the gas cooling rate. So well-fuelled AGN should be more powerful than those with less available gas, but these well-fuelled AGN will necessarily be in higher gas density environments, potentially setting up a balance between AGN feedback and gas cooling sufficient to stop the over-cooling predicted from radiative processes \citep{fabian1994}. Evidence for the co-location of a central AGN with a cold gas fuel supply comes from CO detections in BCGs at the centres of cool-cores that display both radio and optical line emission \citep{edge2001}.   

Previous work has provided tantalising clues on the interrelation between the ICM and the galaxy population. For example looking at the $L_{X}-T_{X}$ relation for clusters that contain radio emission, those that have extended radio structure have a significantly steeper relationship than those that are unresolved \citep{mag2007}. In terms of cool-core clusters, as mentioned above, \cite{mittal2011} find that those with strong cool-cores have a significantly steeper $L_{X}-T_{X}$ than those lacking a cool-core and that the presence of a cool-core indicates a radio-loud central galaxy. There is clearly a relationship between both of these observations, in that it seems that to guarantee a central AGN requires a strong cool-core, and cool-cores and/or AGN lead to a steepening of the $L_{X}-T_{X}$ relation. This demonstrates perhaps that cool-cores can dominate over AGN feedback in the cluster environment, while AGN become more significant in the group regime. Further evidence for this picture is presented in \cite{lin2007} who find that the ratio of energy injected by AGN to the thermal energy of the ICM decreases with increasing cluster mass and recent observations of the radio heating of the X-ray gas around individual early-type galaxies \citep{danielson2012}.

In this paper we perform the first comprehensive study of the interplay between the stellar population of the BCG, its supermassive black hole and the ICM by investigating the optical and radio properties of BCGs within the context of the cluster/group $L_{X}-T_{X}$ relation. The sample used consists of 123 $z<0.3$ clusters taken from the XMM Cluster Survey (XCS, \citealt{romer2001,lloyddavies2011,mehrtens2011}), a serendipitous survey of clusters from the XMM-Newton archive, a large sample which contains a range of systems from the group scale up to massive clusters, which is key for this study. The combination of both optical and radio observations means we can assess both the time averaged (through the assumption of the $M_{\rm BH} - M_{\rm Bulge}$ relation) and instantaneous effect of AGN feedback on the ICM. 
The results of this paper are discussed assuming that all BCGs contain supermassive black holes, obeying the $M_{\rm BH} - M_{\rm Bulge}$ relation, that are potential AGN but whose duty-cycle means that we can observe only a fraction of them to be accreting at any one time. Where appropriate we compare our results to those from the OverWhelmingly Large Simulations (OWLS) run which incorporates AGN feedback \citep{schaye2010,mccarthy2010}. 

We begin the analysis in \S\ref{sec:bcgscale} with BCG - cluster X-ray scaling relations, in order to derive a BCG mass to host dark matter halo mass relation. With the assumption of the $M_{\rm BH} - M_{\rm Bulge}$ relation, this allows us to assess the relative significance of the BCG black hole to that of the entire system. In \S\ref{sec:lilxtx} we study the effect of BCG mass, and thus the time averaged effect of its black hole, on the ICM through the location of the cluster within the $L_{X}-T_{X}$ relation. Under the assumption that black holes also require an effective supply of fuel for accretion, in \S\ref{sec:dyn} we also investigate the effect of the projected distance of the BCG from the peak surface brightness of the ICM on the $L_{X}-T_{X}$ relation. We then move on to investigate the instantaneous effect of AGN in \S\ref{sec:rl}, to look for the key factors affecting the BCG radio-loud fraction and the effect of radio-loudness on the $L_{X}-T_{X}$ relation. Finally, in \S\ref{sec:energy} we use simple energetic arguments to explain our findings.

A Lambda Cold Dark Matter ($\Lambda$CDM) cosmology ($\Omega_{\rm m}=0.27$, $\Omega_{\Lambda}=0.73$, $H_{0}=70$ km\,s$^{-1}$ Mpc$^{-1}$) is used throughout this work.        
    
\section{Sample and Data}    
\label{sec:data}
\begin{figure}
   \centering
\includegraphics[scale=0.5]{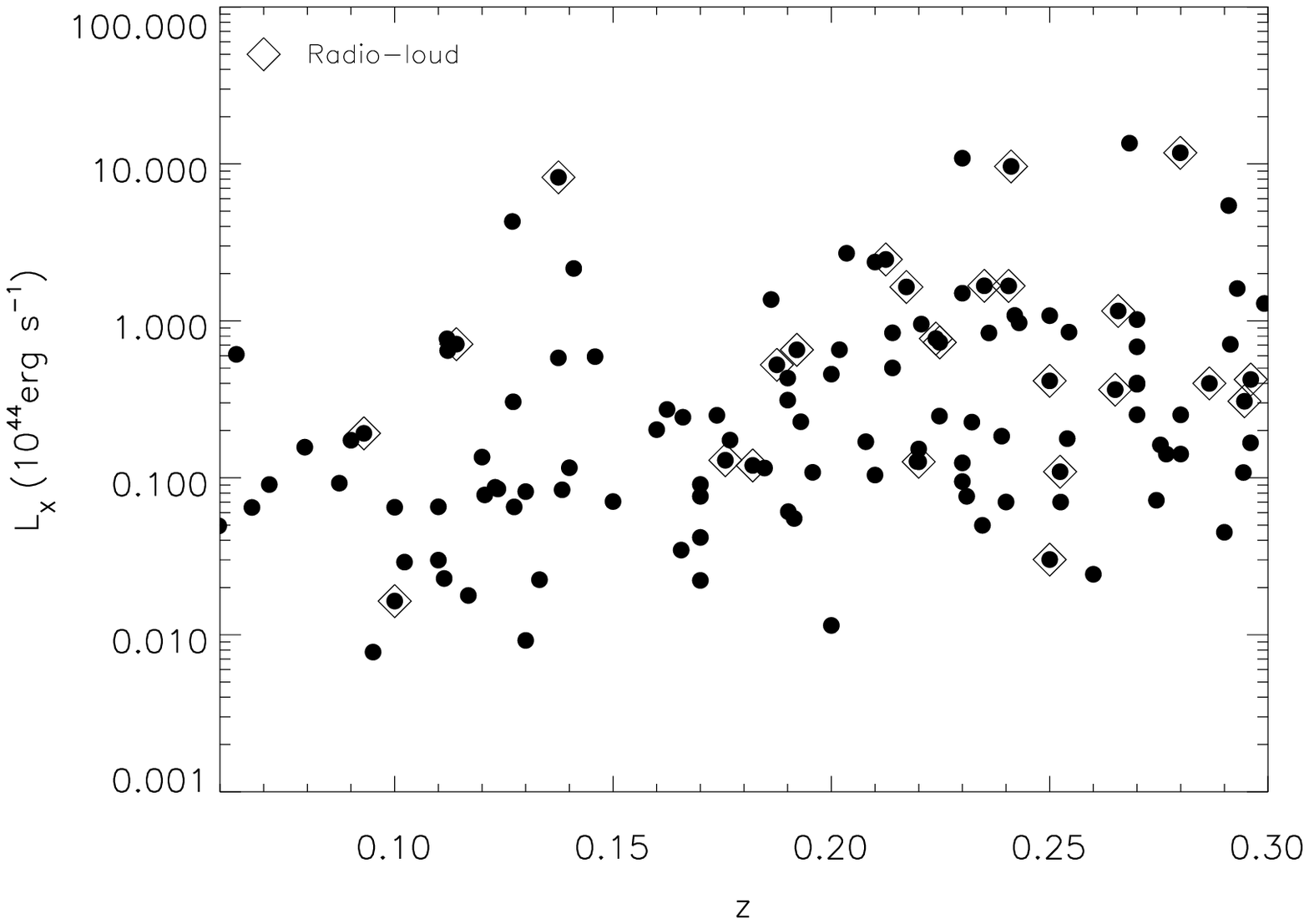} 


\includegraphics[scale=0.5]{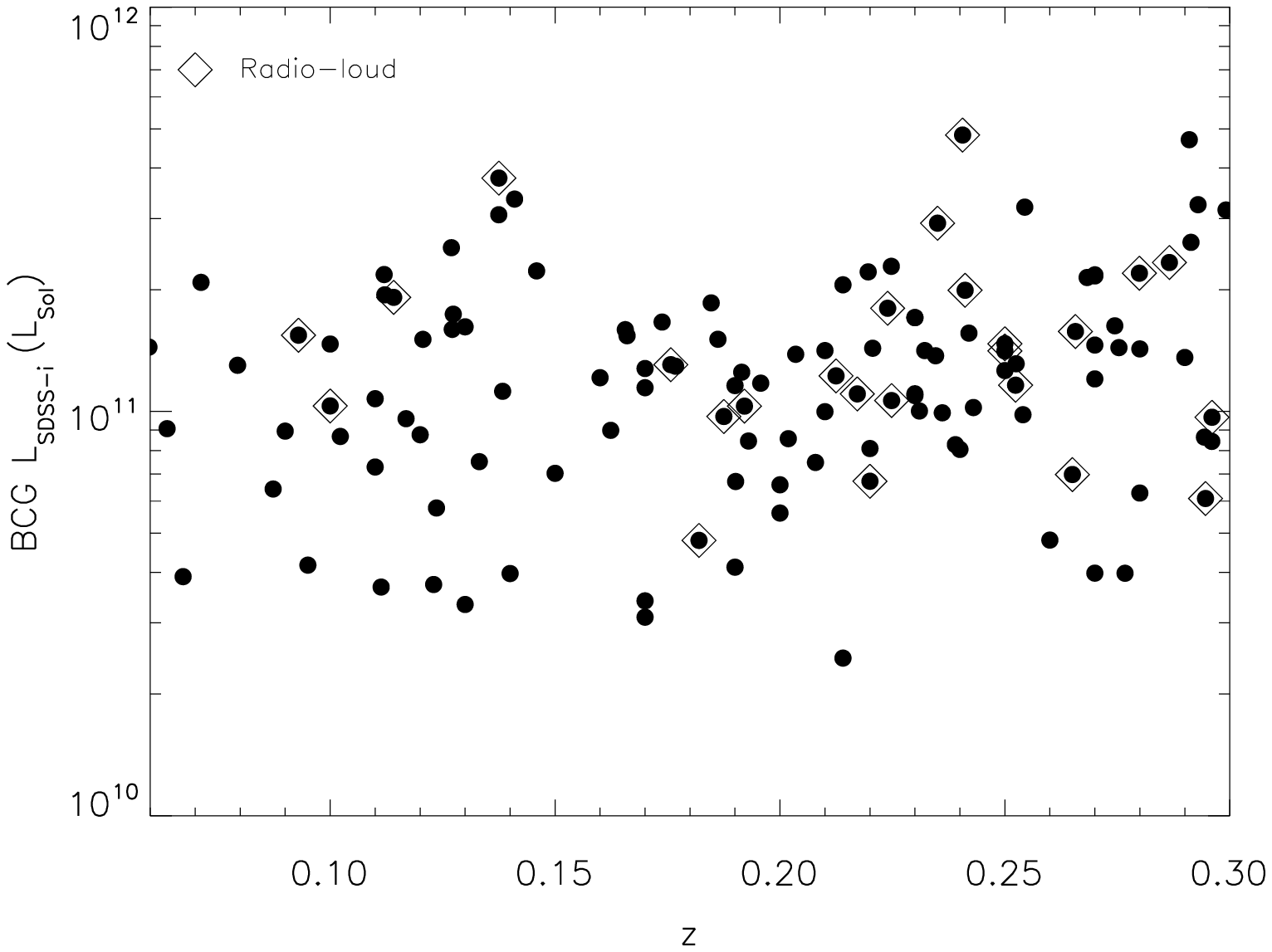} 


\includegraphics[scale=0.5]{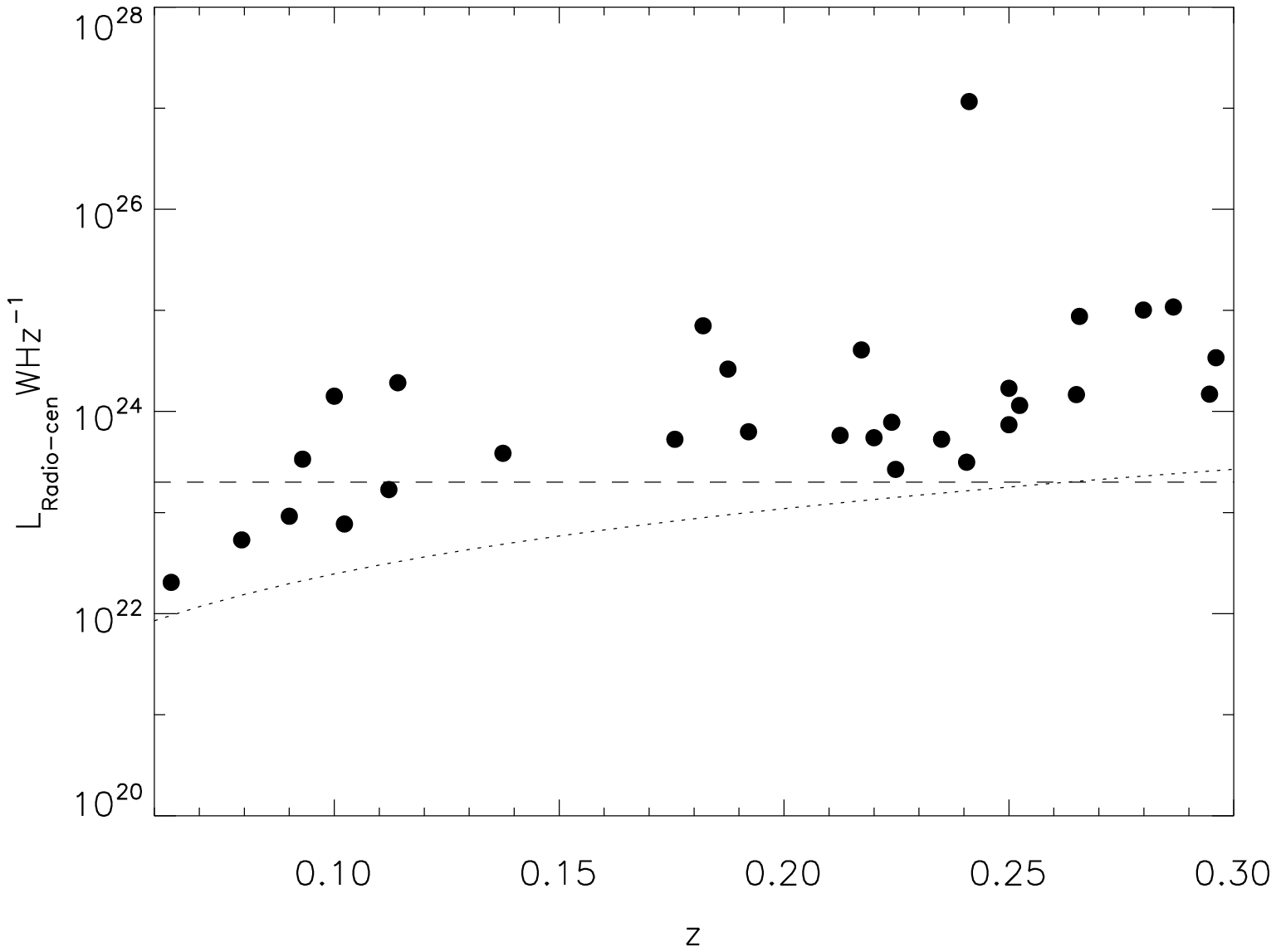} 


\caption{{\it Upper:} The cluster/group bolometric X-ray luminosity plotted against redshift for the XCS-SDSS sample. The radio-loud BCGs are plotted as open diamonds. This demonstrates that $L_{X}$ does not strongly depend on redshift.
{\it Middle:} BCG SDSS $i$-band luminosity plotted against redshift. The radio-loud BCGs are plotted as open diamonds. The sample shows no $L_{i-\rm SDSS}$ - redshift dependence.
{\it Lower:} BCG central radio luminosity ($L_{\rm Radio-cen}$) plotted against redshift for all radio detections in the sample. The horizontal line at $L_{\rm Radio-cen}=2\times10^{23}\rm W\,Hz^{-1}$ represents the cut we introduce to ensure there is no redshift dependent selection in the sample. Throughout this paper we only consider the radio properties of BCGs with an $L_{\rm Radio-cen}$ greater than this (the radio-loud BCGs). The dotted line represents the 1\,mJy threshold of the FIRST catalogue.}
\label{fig:LxLiLrz}
\end{figure}

The galaxy cluster sample and the associated X-ray properties are taken from the XCS first data release \citep{mehrtens2011}, with the data analysis techniques described in \cite{lloyddavies2011} and we therefore point the reader to those papers for the sample details. In order to compare the optical and radio properties of these systems, the additional selection criteria are that the clusters lie within the Sloan Digital Sky Survey Data Release 8 (SDSS DR8) footprint and, where appropriate in this paper, also within the Faint Images of the Radio Sky at Twenty centimetres survey (FIRST survey, \citealt{becker1995}), which has a footprint very similar to SDSS. A redshift cut of $z<0.3$ is also imposed to ensure good optical photometry and completeness of radio detections. The former XCS-SDSS sample consists of 123 clusters and the latter XCS-SDSS-FIRST sub-sample consists of 103 clusters. The XCS has a complicated selection function because of its serendipitous nature and heterogeneous data depths, but Figure \ref{fig:LxLiLrz} ({\it upper panel}) demonstrates there is no evidence for a strong trend between $L_{X}$ and redshift due to a flux limit and, importantly, that we find a similar dynamic range in $L_{X}$ over the entire redshift range considered here. 

The key advantages of the XCS first data release are the sample size and its sensitivity, providing a large range in $L_{X}$ ($\rm0.007 - 14.0\times10^{44} \rm erg\,s^{-1}$, for our sub-sample) and thus an excellent dynamic range in a property related to gas density, essential to probe the effects of AGN on the ICM. The wide range in $T_{X}$ ($0.3 - 10\,\rm keV$, for our sub-sample), and therefore mass, allows us to analyse a significant number of group scale systems, where AGN become more energetically important. Importantly, for a study of clusters that may also harbour X-ray bright AGN, the X-ray data have had any detectable point sources removed, as they were detected within, or in the vicinity of, $\sim25\%$ of the $z<0.3$ sample. These are dealt with using the XAPA vision model described in \cite{lloyddavies2011}, that can detect and remove point sources embedded in extended sources and differentiate between blended point sources and genuine extended sources. 
The X-ray properties and their associated errors can be found in \cite{mehrtens2011} but we illustrate the X-ray errors in the Appendix Figure \ref{fig:lxtxerr}. The X-ray luminosities ($L_{X}$) quoted throughout the paper are bolometric and are calculated within $R_{500}$, the radius at which the cluster mean gravitational mass density drops to $500\times$ that of the critical density. The temperatures are also calculated within $R_{500}$, using spectral fitting carried out with XSPEC and the maximum likelihood Cash statistic \citep{cash1979}. The models used in the fitting include a photoelectric absorption component (WABS; \citealt{morrison1983}) to simulate the nH absorption and a hot plasma component (MEKAL; \citealt{mewe1986}) to simulate the X-ray emission from the ICM (for full details see \citealt{lloyddavies2011}).  

We calculate $M_{500}$ masses for the clusters using an evolving mass - $T_{X}$ scaling relation as described in \cite{sahlen2009} and the virial mass-to-$M_{500}$ conversion from \cite{hu2003}, with the average conversion from virial mass to $M_{500}$ being a factor of $0.79\pm0.02$. The $M-T_{X}$ relation is normalised using HIFLUGCS data for the local cluster population \citep{reiprich2002}. Implicit in the mass determination is also the assumption of an NFW halo profile with a concentration parameter of 5 \citep{navarro1995}. As the redshift range is relatively small, the result of this derived $M_{500} - T_{X}$ scaling is well fit by the relation: 
\begin{equation}
\log\,(M_{500})=1.432 (\pm0.011)\,\log\,(T_{X}) + 13.722 (\pm0.005),
\end{equation}
with $M_{500}$ in $\rm M_{\odot}$ and $T_{X}$ in keV. The XCS sub-sample in this paper covers the mass range $10^{13}<M_{500}<10^{15}$.

The BCG optical properties are extracted from the SDSS DR8 $i$-band model magnitude photometry which assumes a de Vaucouleurs profile fit to obtain a total magnitude. Such a fit is appropriate for the BCGs in our sample as we find the median S\'{e}rsic $n=4.0\pm0.4$ for a sample of X-ray selected clusters at $z\sim0.2$ observed with the {\it Hubble Space Telescope} \citep{stott2011}. The magnitudes are converted to a luminosity ($L_{i-\rm SDSS}$) using $k$ and passive evolution corrections from a \cite{bc03} simple stellar population (SSP) with a Chabrier IMF, a formation redshift, $z_{f}=3$ and solar metallicity, a model appropriate for BCGs \citep{stott2010}. A correction for Galactic extinction using the dust maps of \cite{schlegel1998} is used. We also calculate a stellar mass for the BCGs using an $i$-band mass-to-light ratio of 1.91 derived from the same SSP. Figure \ref{fig:LxLiLrz} ({\it middle panel}) demonstrates that optical incompleteness is not an issue out to $z=0.3$.

The radio data are taken from the FIRST survey catalogue, which has a detection threshold of 1\,mJy. From catalogue matching and visual inspection of the FIRST imaging we obtain the flux from the radio source centred on the BCG and also sum the flux from all radio sources associated with the BCG (i.e. central point source and extended lobes). We note that all of our radio sources have more than just a central component and can therefore all be classified as extended. The 1.4\,GHz radio luminosity of the central and total components ($L_{\rm Radio-cen}$ and $L_{\rm Radio-tot}$) of the BCGs is calculated with a $k$ correction assuming a spectral slope where flux, $S_{\nu}\propto \nu^{-\alpha}$ and $\alpha=0.72$ \citep{coble2007}. To ensure the radio detected fraction of the XCS-SDSS-FIRST sample is complete to $z=0.3$, we plot $L_{\rm Radio-cen}$ against redshift in Figure \ref{fig:LxLiLrz} ({\it lower panel}) and implement a luminosity limit such that $L_{\rm Radio-cen}>2\times10^{23}\rm W\,Hz^{-1}$ to account for the flux limited nature of the sample. Hereafter $L_{\rm Radio-cen}>2\times10^{23}\rm W\,Hz^{-1}$ defines the `radio-loud' sample, with the remaining XCS-SDSS-FIRST clusters being `radio-quiet'.  We note, that this is similar to the lower limit used in \cite{best2007} and \cite{lin2007} ($1\times10^{23}\rm W\,Hz^{-1}$) with the former finding that the contamination of the radio-loud BCG sample with star forming galaxies is minimised to $\lesssim1\%$ using this criterion. The most luminous source in the sample is the cluster associated with the well-studied radio galaxy 4C +55.16 with $L_{\rm Radio-tot}=1.4\times10^{27}\rm W\,Hz^{-1}$ at $z=0.24$.

The BCG selection for a cluster is usually obvious from visual inspection of the images as they are the prominent galaxy closest to the X-ray centroid and often have a cD-like profile. However we choose to formalise this by studying the tip of the red sequence in the colour magnitude relation. For each cluster we identified the red sequence with $g - r$ colour and selected the brightest galaxy from the $r$-band magnitudes of all the red sequence galaxies within a projected distance of 500\,kpc from the cluster X-ray centroid as for approximately 95\% of clusters the BCG lies within this radius \citep{lin2004}. 

A table of optical and radio data from this paper can be found at: {\it http://www.astro.dur.ac.uk/~jps/Stott2012/Stott2012.cat}


\subsection{OWLS Comparison simulations}
\label{sec:owls}
For comparison with theory we compare the results in this paper with simulations from the OverWhelmingly Large Simulations project (OWLS, \citealt{schaye2010, mccarthy2010}) that are ideal for this study as they include the relevant physics, and output values for cluster mass, $L_{X}$, $T_{X}$, galaxy stellar mass and photometry. The simulated $T_{X}$ we choose for comparison with our observations is emission weighted rather than spectroscopic-like, although for OWLS the difference between these is small with a median offset of 0.1\,keV. The volume-limited nature of the simulations means that they do not contain the massive clusters found in the XCS sample, but there is a significant crossover regime between the two ($10^{13}\lesssim M_{500}\lesssim10^{14} \rm M_{\odot}$). The two OWLS simulations we use are both high resolution, hydrodynamical, cosmological simulations that include radiative cooling \citep{wiersma2009a}, star formation \citep{schaye2008}, feedback from supernovae driven winds \citep{DallaVecchia2008} and full chemodynamics \citep{wiersma2009b}, but only one of the simulations includes feedback from supermassive black holes (AGN feedback, \citealt{booth2009}). While the efficiency of the AGN feedback was set to reproduce the normalisation of the $M - \sigma$ relation, the model was not tuned to reproduce any other observables. As demonstrated by \cite{mccarthy2010, mccarthy2011}, in the AGN feedback simulation the low entropy gas, which would otherwise condense to form stars, is removed or heated at high redshift ($z>1$) by energy input from AGN in typical $L^{\star}$ galaxies. This needs to happen at high redshift as the gravitational binding energies are easier to overcome in the lower mass progenitors of the $z=0$ groups. Subsequent to the ejection of the low entropy gas, the injection of further energy by the AGN of the central galaxy stops the remaining gas from over-cooling in the centre of the group and thus stops the gas from causing significant star formation.

\begin{figure}
   \centering
\includegraphics[scale=0.5]{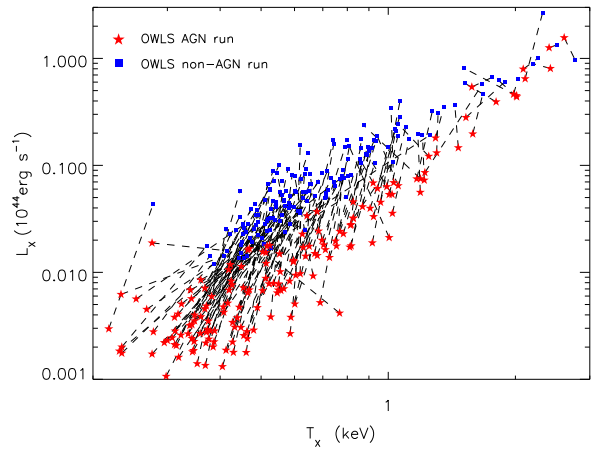} 
\caption[]{A plot showing the effect of AGN feedback on the X-ray luminosity - temperature relation in the OWLS simulation (as demonstrated in \citealt{mccarthy2010}). To help illustrate this point the temperatures here are cool-core corrected. Blue squares are the non-AGN feedback run and red stars are with AGN feedback included. The connecting lines show where each individual cluster is moved to by the feedback process. The low entropy gas, which would otherwise condense to form stars, is removed or heated at high redshift ($z>1$) by energy input from AGN in typical $L^{\star}$ galaxies. The ejection of low entropy gas and the injection of energy by the AGN in the central galaxy stops the remaining gas from over-cooling in the centre of the group and thus stops the gas from causing significant star formation in the central galaxy at late times.}
   \label{fig:owls}
\end{figure}

Figure \ref{fig:owls} illustrates the effect of this AGN feedback on the $L_{X} - T_{X}$ relation within the OWLS simulations, as was also demonstrated in \cite{mccarthy2010}. Here, the lines on the plot connect the same halos from the non-AGN feedback run to those with AGN feedback included. We see that when comparing the AGN and non-AGN runs the effect of the AGN, to first order, is to lower the $L_{X}$ for a given $T_{X}$. This is because the simulated AGN are heating and removing the low entropy gas, lowering the central density and thus $L_{X}$ which depends on the integral of the square of the gas density. The influence of the AGN is more pronounced in the lowest $T_{X}$, and therefore lowest mass, groups as these systems have shallower gravitational potential wells and are therefore less able to hold onto their central gas. Similar predictions for the $L_{X} - T_{X}$ relation have been made by other groups (e.g. \citealt{puchwein2008,fabjan2010}). 

As in \cite{mccarthy2010, mccarthy2011}, groups/clusters are selected on the basis of halo mass. Haloes are identified using a friends-of-friends (FOF) algorithm and we select only those haloes with $M_{200} > 10^{13} \rm M_{\odot}$. Self-gravitating subhaloes are identified in each FOF system using the SUBFIND algorithm of \cite{dolag2009}, which is modified version of that originally developed by \cite{springel2001}.  The BCG is defined as the stellar component associated with the most massive subhalo in a FOF system.  Since our observations do not include ICL, we only quote the total stellar mass/luminosity of the BCG within a radius $< 30 \,$kpc. We find that the AGN feedback in the OWLS simulation reduces the stellar mass of the BCG by a factor of $\sim20$ to bring them into agreement with values derived from observation. As a consistency check, we compare our derived mass-to-light ratio (1.91) used in \S\ref{sec:data} to obtain the stellar masses for our observed BCGs and find the median value of that in the simulations to be in excellent agreement at $1.95\pm0.03$. The OWLS output data, presented as comparison throughout, is at $z=0$.

\section{Results}

\subsection{BCG X-ray and mass scaling relations}
\label{sec:bcgscale}
The optical properties of BCGs are known to correlate with the X-ray properties of their host clusters \citep{edge1991,collins1998,lin2004,popesso2007,whiley2008,stott2008,brough2008,mittal2009}. These relations have been studied for large samples of massive clusters but, owing to the wide mass range of the XCS, we can now probe them for a consistently analysed sample of groups and clusters, spanning over three decades in $L_{X}$. A summary of the scaling relations is presented in Table \ref{tab:scaling}.  

\begin{figure}
   \centering
\includegraphics[scale=0.5]{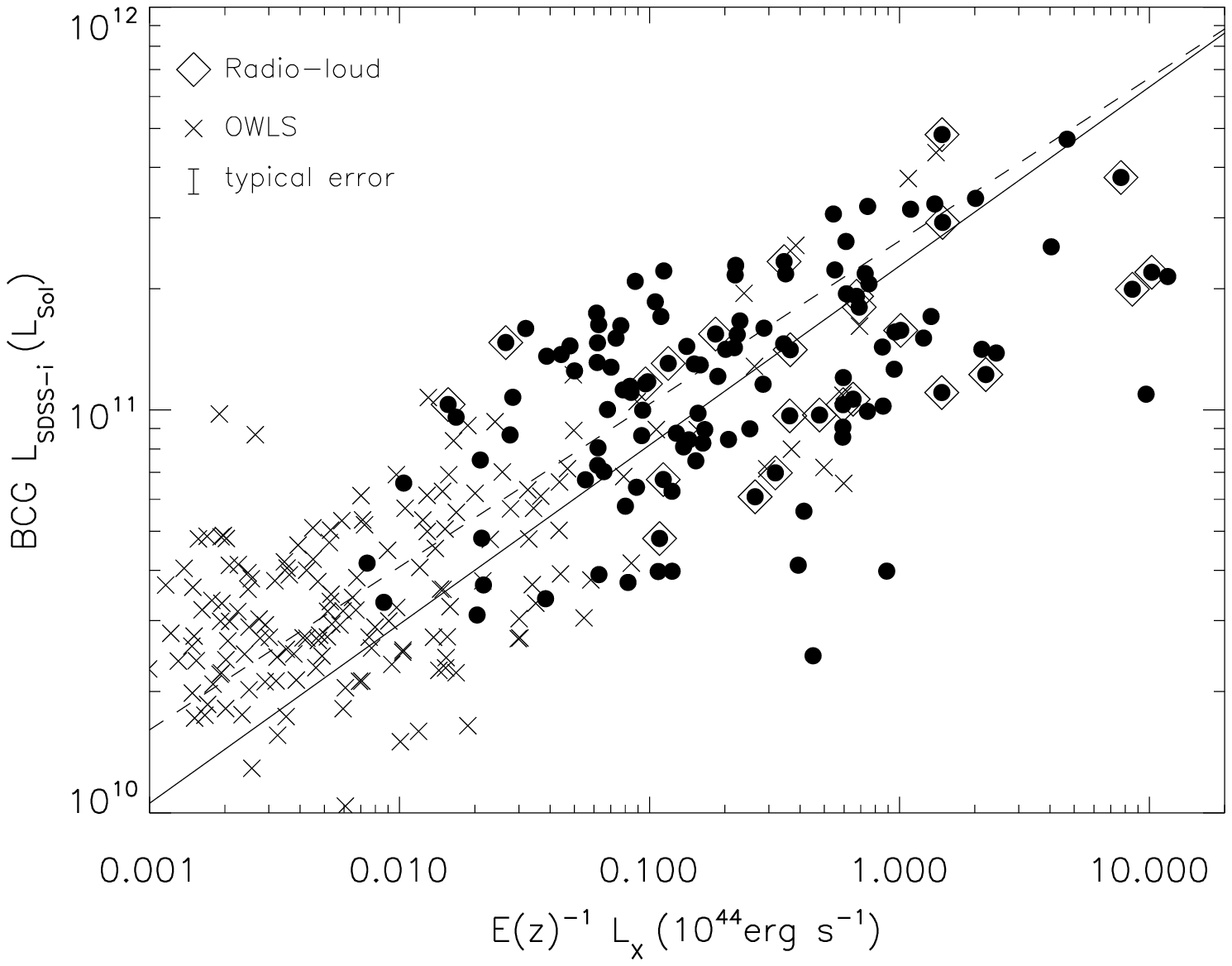} 


\includegraphics[scale=0.5]{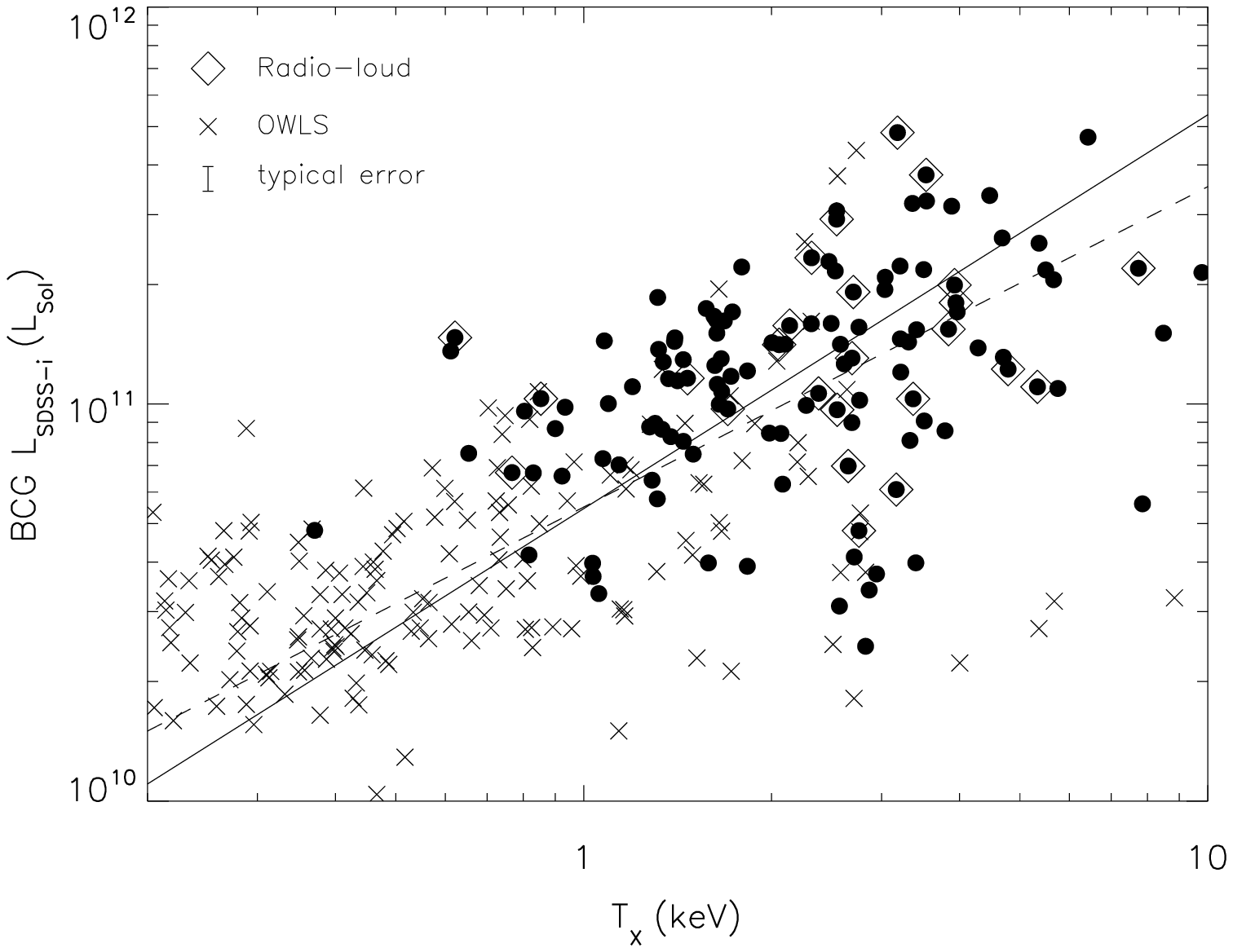} 


\includegraphics[scale=0.5]{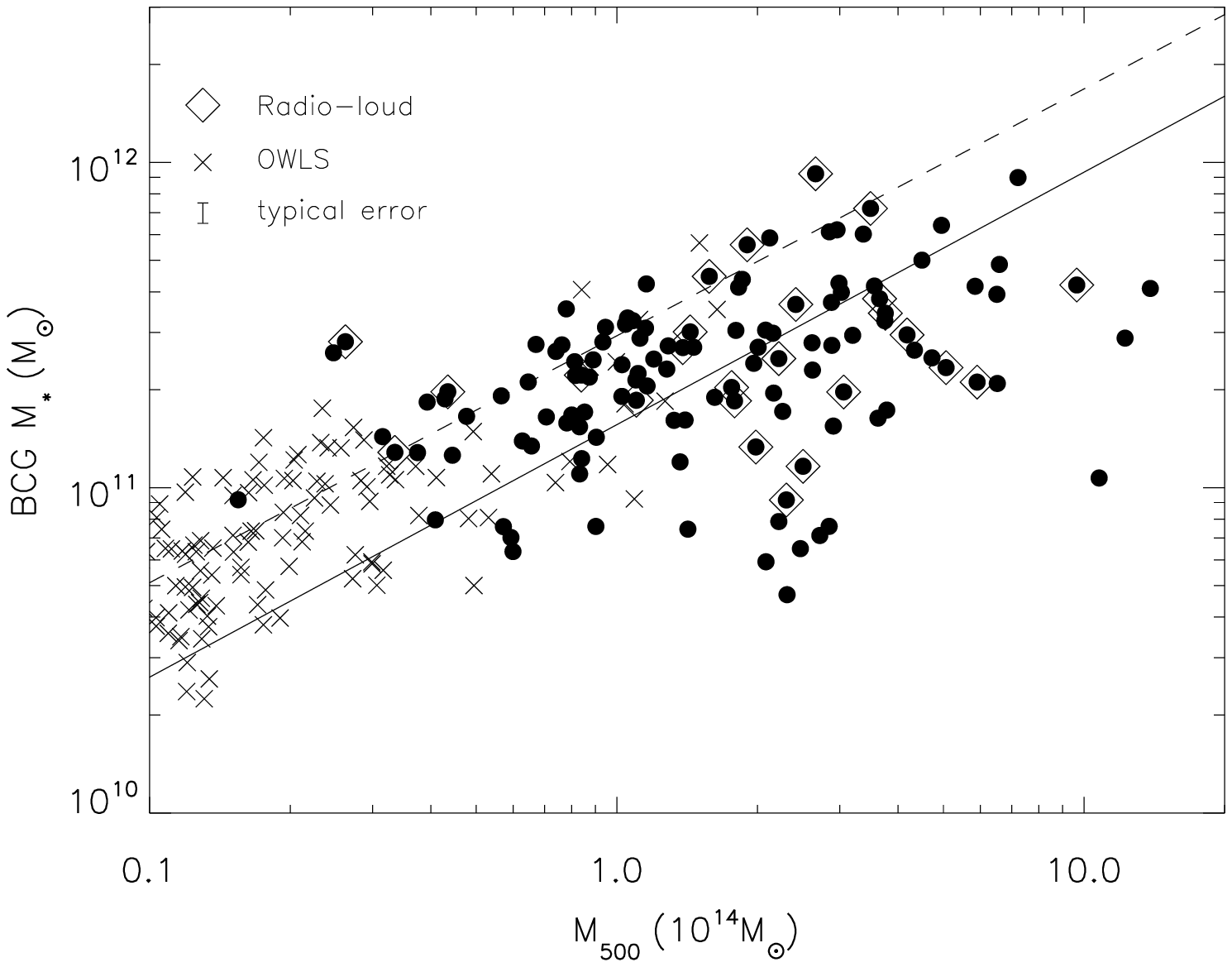} 

\caption{For the BCG - cluster scaling relation plots presented here, the observed and simulated data are represented as filled points and crosses respectively, and fits to the former and latter are solid and dashed lines. The radio-loud BCGs are plotted as open diamonds. The error bars for the X-ray data can be found in Figure \ref{fig:lxtxerr}. We note that the {\it middle} and {\it lower} panels are not independent - see \S\ref{sec:data} for a description of how $M_{500}$ was calculated. {\it Upper:} BCG SDSS $i$-band luminosity plotted against cluster/group bolometric X-ray luminosity for the XCS-SDSS sample. 
{\it Middle:} BCG SDSS $i$-band luminosity plotted against cluster/group X-ray temperature for the XCS-SDSS sample.
{\it Lower:} BCG stellar mass plotted against cluster mass.}
\label{fig:bcgscale}
\end{figure}

Figure \ref{fig:LxLiLrz} demonstrates that there is no evidence for a strong correlation in either $L_{X}$ or $L_{i-\rm SDSS}$ with redshift due to either an X-ray or optical flux limit (which could perhaps be present if cluster confirmation relied on bright BCGs). Satisfied that any correlations will not be due to a selection effect, in Figure \ref{fig:bcgscale} ({\it upper and middle panels}) we plot the $L_{i-\rm SDSS}$ of the BCG against the $L_{X}$ and $T_{X}$ of the host clusters. This is the first time such relations have been produced for a large sample of BCGs over such a wide range in $L_{X}$ and $T_{X}$. There is clearly a positive relationship between BCG $L_{i-\rm SDSS}$ and both $L_{X}$ and $T_{X}$. We quantify this by performing a linear regression, BCES bisector fit \citep{isobe1990,akritas1996} to the logged data, accounting for the errors in both quantities and the intrinsic scatter, finding that: 
\begin{equation}
\log\,(L_{i-\rm SDSS}) = (0.44\pm0.04)\,\log\,(E(z)^{-1}\, L_{X}) + (11.36\pm0.03)
\end{equation}
\begin{equation}
\log\,(L_{i-\rm SDSS}) = (0.99\pm0.05)\,\log\,(T_{X}) + (10.74\pm0.03),
\end{equation}
where $L_{i-\rm SDSS}$ is in units of $L_{\odot}$, $L_{X}$ is in units of $10^{44}\rm \rm erg\,s^{-1}$, $T_{X}$ is in $\rm keV$ and $E(z)=[\Omega_{\rm m}(1+z)^3 + \Omega_{\Lambda}]^{1/2}$. 

We compare these findings with the OWLS AGN feedback run, plotting its output in the panels of Figure \ref{fig:bcgscale} (crosses). In general the OWLS simulation covers a lower cluster mass ($L_{X}$ and $T_{X}$) range than the observations due to the volume-limited nature of the simulations. As discussed in \S\ref{sec:owls}, the simulated BCG photometry and stellar mass give an average mass-to-light ratio consistent with that of our observed photometry and SSP derived masses. The 30\,kpc radius aperture used to extract the simulated magnitudes is appropriate for comparison with the SDSS model magnitudes as BCGs typically have effective radii of 30-40\,kpc \citep{stott2011}. The plots show that the OWLS simulations are in good agreement with both of the observed X-ray -- BCG luminosity data, as where there is an overlap, the $L_{i-\rm SDSS}$ values agree and if taken together the two samples appear to form a continuous distribution. To quantify this we also perform fits to the OWLS data which yield BCG $L_{i-\rm OWLS}\propto L_{X-\rm OWLS}^{0.40\pm0.02}$ and BCG $L_{i-\rm OWLS}\propto T_{X-\rm OWLS}^{0.81\pm0.04}$, with the former in excellent agreement with its observational counterpart. 


The relations seen in Figure \ref{fig:bcgscale} ({\it upper and middle panels}) demonstrate that BCG stellar mass correlates with the total mass of the cluster. To quantify this, we study the derived BCG stellar masses and cluster $M_{500}$ values calculated in \S\ref{sec:data} and compared to the OWLS output. In Figure \ref{fig:bcgscale} ({\it lower panel}) we plot BCG stellar mass against the cluster $M_{500}$ value and perform fits to the observations and simulations. For the observations we find that the BCG stellar mass: 
\begin{equation}
\log\,(M_{\rm *BCG})= (0.78\pm0.06)\,\log\,(M_{500}) + (11.19\pm0.06), 
\end{equation}
with $M_{500}$ in units of $10^{14}\rm M_{\odot}$. This is important as in a cosmological context it gives a relation between the stellar mass of the central galaxy and its dark matter halo mass over the range $10^{13}<M_{500}<10^{15}\rm \rm M_{\odot}$. 

For the OWLS simulations $M_{\rm *BCG-OWLS}\propto M_{500- \rm OWLS}^{0.76\pm0.04}$, again in excellent agreement with observation. The offset between the fit zero-points is most likely due to the uncertainties involved in both deriving physical quantities from observables such as $L_{i-\rm SDSS}$ and $T_{X}$ and the difficulty in simulating realistic baryon physics.

We compare our BCG cluster mass scaling relation to that found in other studies. \cite{haarsma2010} find that studies with photometry calculated in fixed metric apertures typically observe shallower relations between BCG luminosity/mass and cluster mass than those that use isophotal magnitudes, with those using S\'{e}rsic profile derived masses, as is the case in this study, found to be the steepest \citep{mittal2009}. However, we also note that the large fixed 30\,kpc radius aperture from the OWLS simulation does not appear to give a shallower slope. In the comparisons that follow we assume a linear relationship between BCG luminosity and BCG mass for the studies that only provide the dependency of BCG luminosity on cluster mass. The exponent $\alpha=0.78\pm0.06$ in the relation $M_{\rm *BCG}\propto M_{\rm cluster}^{\alpha}$ found in this study is indeed similar to the $\alpha=0.62\pm0.05$ found by extrapolating S\'{e}rsic profiles in \cite{mittal2009} (who also use a BCES fitting routine). For the isophotal and fixed metric aperture studies $\alpha$ is significantly less (e.g. \citealt{lin2004} find $\alpha=0.26\pm0.04$, \citealt{popesso2007} find $\alpha=0.25$, \citealt{whiley2008} find $\alpha=0.12\pm0.03$ and \citealt{brough2008} find $\alpha=0.11\pm0.10$). The reason for this may be that the S\'{e}rsic-like photometry gives brighter total magnitudes and perhaps therefore raise the value of $\alpha$.

BCGs can be described as galaxies at the centres of their dark matter halos and thus we compare to mass scaling relations derived from clustering and halo occupation modelling. For example, \cite{yang2005} find $L_{*}\propto M_{\rm halo}^{\alpha}$ where $\alpha=0.25$ for halo masses above $10^{13}h^{-1}\rm \rm M_{\odot}$ when studying the halo occupation statistics of central galaxies in groups and $\alpha=0.28$ inferred by \cite{vale2004} from matching the galaxy luminosity function to a theoretical halo mass function. It is clear that our $\alpha$ is significantly higher than these alternatively derived values, however, \cite{zheng2009} find $\alpha\sim0.5$ from the clustering of central luminous red galaxies (LRGs) at $z=0.3$ which is closer to that found here and a more appropriate comparison sample.

\begin{table*}
\begin{center}
\caption[]{BCG - Cluster scaling relation summary of the form $\log\,Y=a\,\log\,X+b$.}
\label{tab:scaling}
\small\begin{tabular}{llll}
\hline
$Y$ & $X$ & $a$  & $b$\\
\hline
BCG $L_{i-\rm SDSS} (L_{\odot})$&$T_{X} (\rm keV)$						&0.99$\pm$0.05&10.74$\pm$0.03\\
BCG $L_{i-\rm SDSS} (L_{\odot})$&$L_{X} (10^{44} \rm erg\,s^{-1})$		&0.44$\pm$0.04&11.36$\pm$0.03\\
OWLS BCG $L_{i} (L_{\odot})$&$T_{X} (\rm keV)$						&0.81$\pm$0.04&10.74$\pm$0.03\\
OWLS BCG $L_{i} (L_{\odot})$&$L_{X} (10^{44} \rm erg\,s^{-1})$		&0.40$\pm$0.02&11.42$\pm$0.05\\
$M_{\rm *BCG} (\rm \rm M_{\odot})$&$M_{500} (\rm \rm 10^{14} M_{\odot})$			&0.78$\pm$0.06&11.19$\pm$0.06\\
OWLS $M_{\rm *BCG} (\rm \rm M_{\odot})$&$M_{500} (\rm \rm  10^{14} M_{\odot})$		&0.76$\pm$0.04&11.47$\pm$0.04\\

\hline
\end{tabular}
\end{center}
\end{table*}

\subsubsection{BCG colours}
\begin{figure}
   \centering
\includegraphics[scale=0.5]{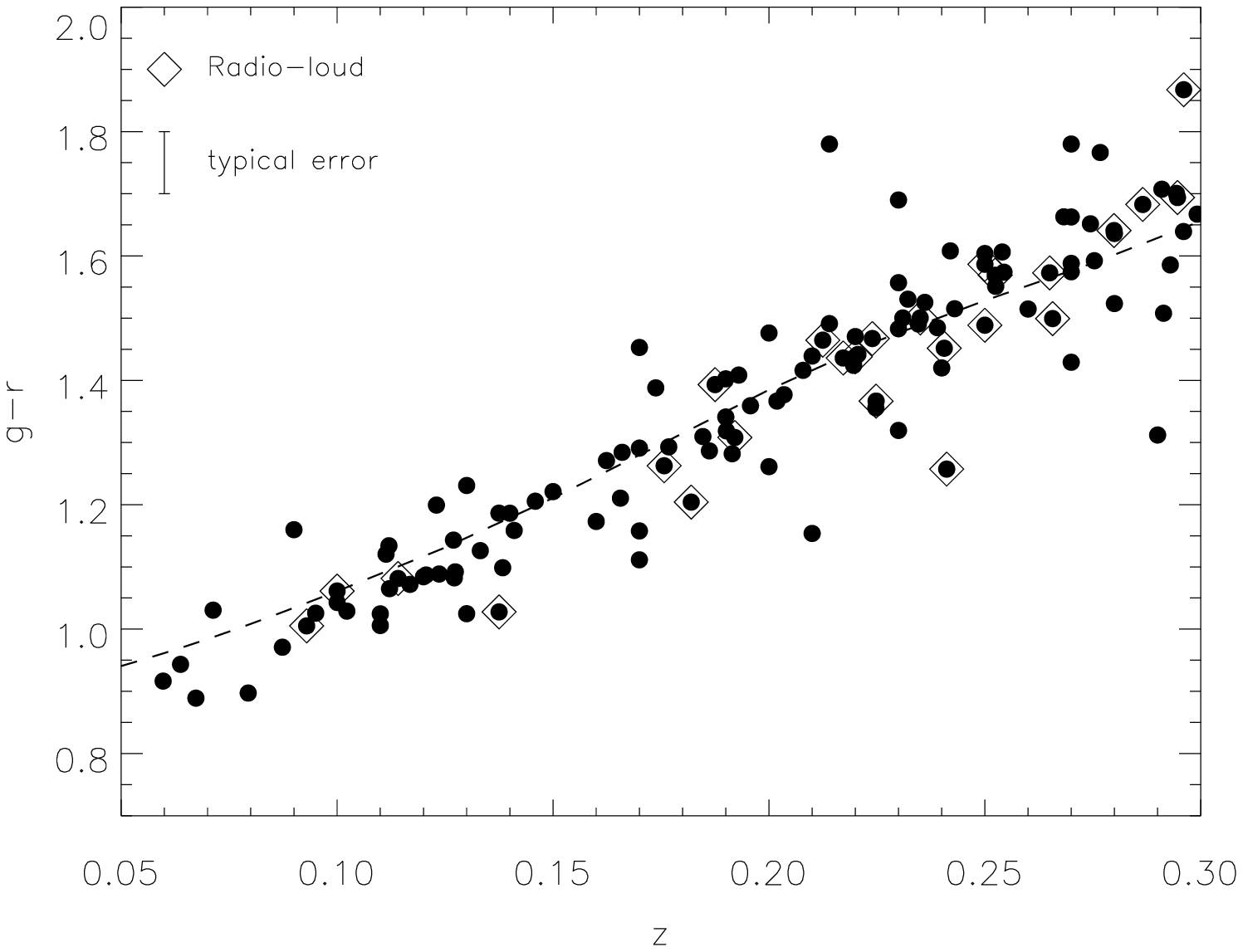} 

\caption[]{BCG SDSS $g-r$ colour plotted against redshift for the XCS-SDSS sample (filled points). A  \cite{bc03} simple stellar population (SSP) with a Chabrier IMF, a formation redshift, $z_{f}=3$ and solar metallicity is plotted as the dashed line. The radio-loud BCGs are plotted as open diamonds.}
   \label{fig:col}
\end{figure}

We also investigate BCG colours to look for those that are especially blue or red by seeing how the $g-r$ colour evolves with redshift. Blue colours may indicate star formation, perhaps linked to cool-core activity \citep{odea2008}. Instead, we find that BCGs are remarkably uniform in colour with an r.m.s scatter of 0.09\,mag across the entire redshift range, which is consistent with the photometric uncertainty. The colour evolution is in excellent agreement with the \cite{bc03} SSP used throughout the paper (Figure \ref{fig:col}) demonstrating that the stellar populations in BCGs are consistent with being formed at $z=3$. We note that there is no correlation between colour and any other BCG or cluster property discussed throughout this study.

\subsection{The effect of BCG mass on the $L_{X} - T_{X}$ relation}
\label{sec:lilxtx}
\begin{figure*}
   \centering
\includegraphics[scale=0.8]{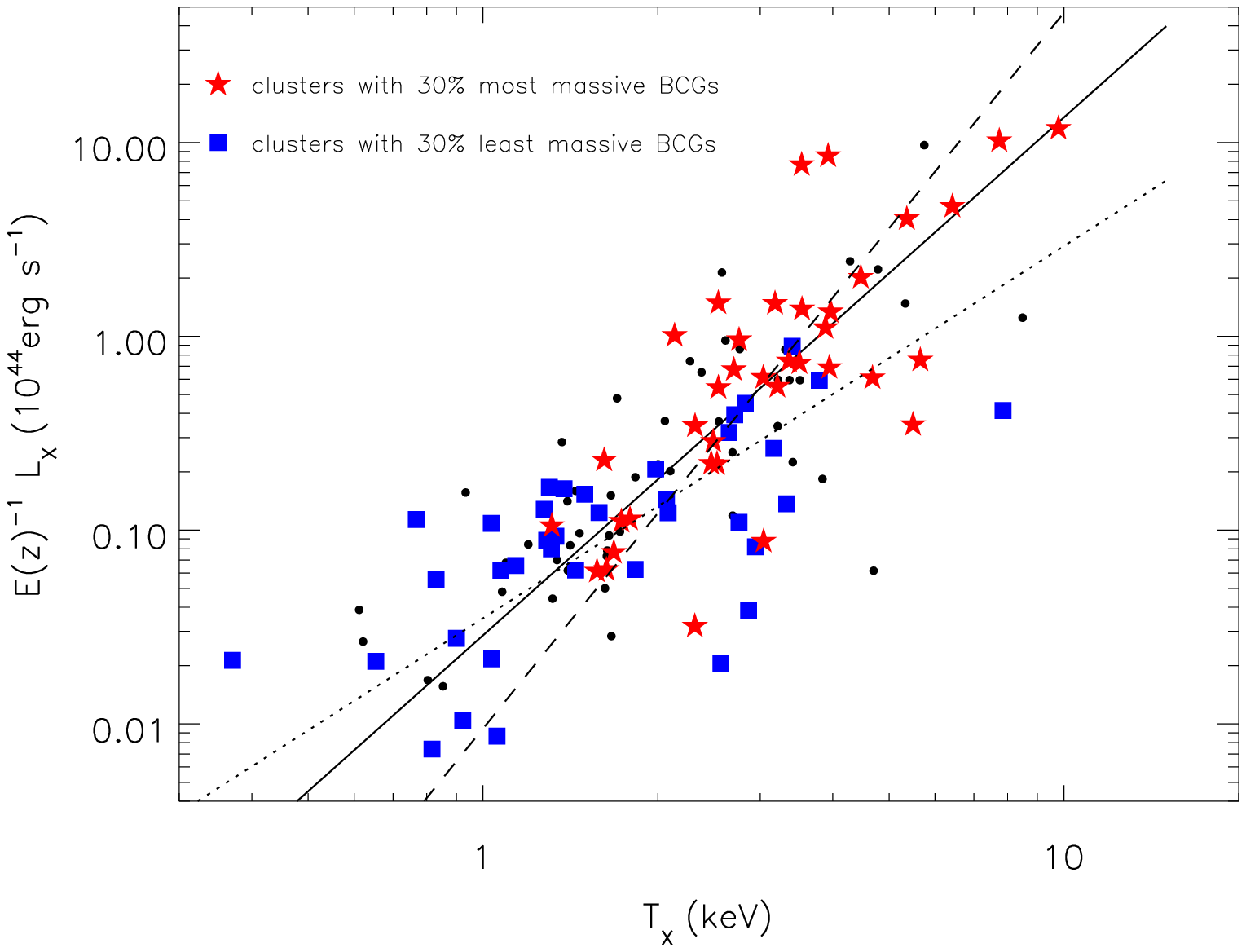} 

\caption{The X-ray luminosity plotted against X-ray temperature. Red stars are the top 30\% of clusters in terms of BCG SDSS i-band luminosity and blue squares are the bottom 30\%. The dashed line is fit to the clusters with the most luminous BCGs, the dotted line is fit to those with the least luminous and the solid line is fit to all (see also Hilton et al. in prep). The luminous population is found to have a steeper slope than the faint population. The error bars for the X-ray data can be found in Figure \ref{fig:lxtxerr}.}
   \label{fig:lxtx}
\end{figure*}

As discussed in the introduction, the $L_{X} - T_{X}$ relationship is used to understand the physics of the ICM with, to first order, $L_{X}$ correlating with the integral of the gas density squared and $T_{X}$ correlating with the virial mass of the system. In Figure \ref{fig:lxtx} we plot the $L_{X} - T_{X}$ relation for the groups and clusters in the XCS-SDSS sample. Indicated on the plot are the most luminous (red stars) and the least luminous (blue squares) 30\% of BCGs (37 clusters in each). A linear regression fit to the logged $L_{X} - T_{X}$ data, accounting for the errors in both quantities (see \citealt{mehrtens2011} for error values) and the intrinsic scatter, is performed for these two subsamples and the whole sample assuming the form $\log\,(E(z)^{-1}\,L_{X})=a\,\log\,(T_{X}) + b$, with $L_{X}$ in units of $10^{44}\rm erg\,s^{-1}$ and $T_{X}$ in $\rm keV$. As in \S\ref{sec:bcgscale} the fitting routine used is the BCES bisector method with errors derived from bootstrapping \citep{isobe1990,akritas1996}. We note that the results are not affected if we instead use the BCES orthogonal method which gives fit parameters consistent to within the $1\sigma$ error of the preferred bisector method. The XCS selection function is not taken into account in these fits; however, there is little redshift dependence in the $L_{X}$ limit over the redshift range considered (Figure \ref{fig:LxLiLrz}, {\it upper panel}) and so we assume selection effects will not have a significant impact on our results (although see \S\ref{sec:bcgopt} for a discussion on Malmquist bias). For the whole XCS-SDSS sample these best-fit parameters are $a=2.67\pm0.19$ and $b=-1.54\pm0.07$ (see also Hilton et al. in prep)\footnote{Hilton et al. (in prep) will present the XCS $L_{X} - T_{X}$ relation for the entire sample and its evolution with redshift.}. This is typical of group and cluster samples with $a>2$ indicating a break from self-similarity \citep{mushotzky1984,edge1991b,markevitch1998,arnaud1999,pratt2009,mittal2011}.

For the luminous BCG $L_{i-\rm SDSS}$ sub-sample these parameters are $a_{\rm lum-30}=3.69\pm0.49$ and $b_{\rm lum-30}=-2.02\pm0.24$ and for the faint BCG $L_{i-\rm SDSS}$ sub-sample they are $a_{\rm faint-30}=1.92\pm0.33$ and $b_{\rm faint-30}=-1.45\pm0.11$. The $L_{X} - T_{X}$ relation for the most massive BCGs is steeper than for the lowest mass BCGs to a significance of $3\sigma$, with a cross-over at $T_{X}\sim2\,\rm keV$. There is therefore good evidence that above $T_{X}\sim2 \rm \,keV$, the most massive BCGs live in clusters with a higher $L_{X}$ value for a given $T_{X}$ than those that host the least massive BCGs, with the fits to the two populations crossing at $T_{X}\sim2\rm \,keV$. Clusters that host the least massive BCGs seem to follow a relation in agreement with the self-similar, $a=2$, case. These results are summarised in Table \ref{tab:lxtx}. 

This steepening of the $L_{X} - T_{X}$ relation is also found in other sub-populations, with \cite{mittal2009} finding strong cool-core clusters have $L_{X}\propto T_{X}^{3.33\pm0.15} $ and \cite{mag2007} finding $L_{X}\propto T_{X}^{4.0\pm0.2}$ for those containing BCGs with extended radio emission. Both these properties may be associated with clusters that contain the most massive BCGs, as we discuss in \S\ref{sec:disc}.

\begin{figure}
   \centering
\includegraphics[scale=0.5]{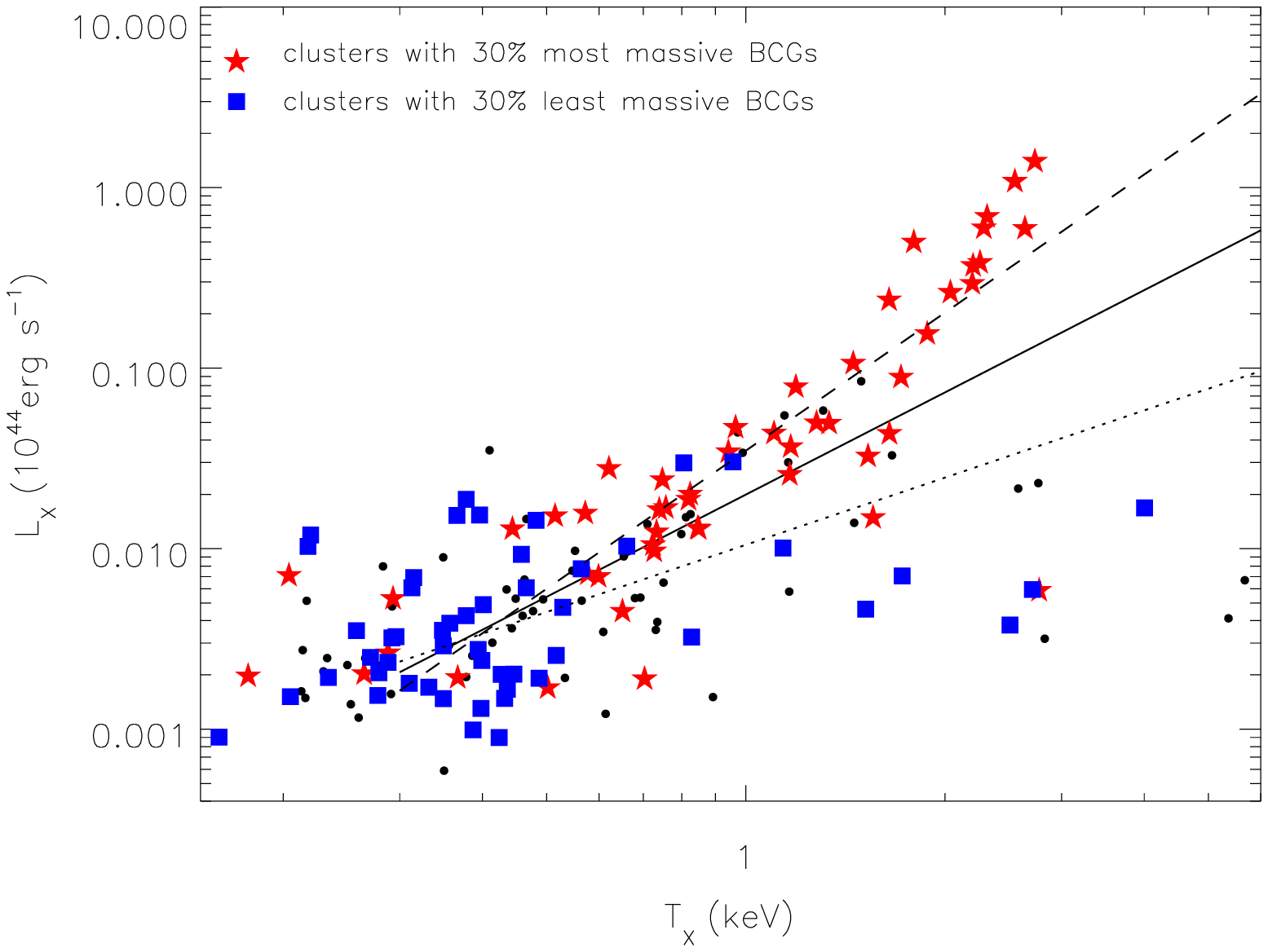} 
\caption{The X-ray luminosity plotted against X-ray temperature for the OWLS simulation. Red stars are the top 30\% of clusters in terms of BCG i-band luminosity and blue squares are the bottom 30\%. The dashed line is fit to clusters with the most luminous BCGs, the dotted line is fit to those with the least luminous BCGs and the solid line is fit to all.}
   \label{fig:owlslxtx}
\end{figure}

For a theoretical point of view, we plot the $L_{X}-T_{X}$ relation for the OWLS simulation and find a qualitatively similar situation, albeit over a different range in $T_{X}$/cluster mass owing to the limited volume of the simulation (see Figure \ref{fig:owlslxtx}). With $a$ and $b$ representing the same fit parameters as above, 
$a_{\rm OWLS}=1.88\pm0.16$ and $b_{\rm OWLS}=-1.70\pm0.07$, $a_{\rm OWLS-lum-30}=2.54\pm0.21$ and $b_{\rm OWLS-lum-30}=-1.45\pm0.06$, and $a_{\rm OWLS-faint-30}=1.24\pm0.13$ and $b_{\rm OWLS-faint-30}=-1.98\pm0.08$. Although the slopes are systematically shallower, these fits still show the same relative behaviour, analogous to the observations.

\subsection{Dynamical state of the cluster}
\label{sec:dyn}
One way of testing whether a cluster is relaxed, is to investigate the projected offset of the BCG from the X-ray centroid of the cluster in units of $R_{500}$, the idea being that in the least disturbed, most virialised systems one would expect both the peak in the gas density and the BCG to be co-located at the centre of mass of the cluster \citep{sanderson2009}. To measure this offset correctly we note that the XCS and SDSS astrometry are matched \citep{mehrtens2011} and that the optical centroiding of the BCG from SDSS is known to $\sim1-2\,\rm kpc$. However, the XCS is a serendipitous survey and as such many of the sources are off-axis and affected by the strongly varying point spread function (PSF) over the {\it XMM Newton} field of view, which has to be corrected for \citep{lloyddavies2011}. We therefore expect the main source of error will be the centroiding of the extended X-ray emission found by the XCS XAPA software. Because the XCS is serendipitous we can make use of the fact that a number of our sources are multiply imaged, some even being the on-axis main target in one exposure but off-axis in others. There are 13 such multiply imaged clusters in the redshift range of our sample and we report no correlation between off-axis angle and centroid offset, so this is clearly not an issue. However, there is a scatter in the centroid positions of the same clusters when measured from different exposures, which has a median offset of 3.3 arcsec, corresponding to an average offset for the sample of 11\,kpc or $\sim0.01\,R_{500}$.

\begin{figure}
   \centering
\includegraphics[scale=0.5]{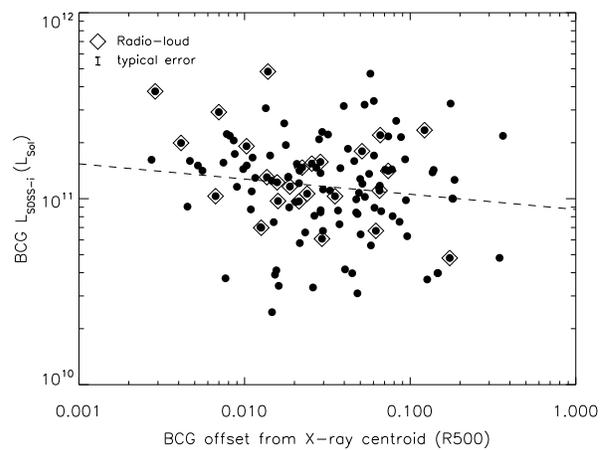} 

\caption{BCG SDSS $i$-band luminosity plotted against BCG offset from X-ray centroid. A fit to these data is represented with a dashed line which shows no significant correlation. The radio-loud BCGs are plotted as open diamonds. }
   \label{fig:lidist}
\end{figure}

The median offset between the BCG and the X-ray centroid for the XCS-SDSS sample presented here is found to be $0.030\pm0.010 \,R_{500}$. Figure \ref{fig:lidist} shows that there is no significant correlation between this offset and the luminosity of the BCG ($L_{i-\rm SDSS} \propto \rm BCG offset^{-0.09\pm0.05}$). This is surprising as one might expect that BCGs at the centre of clusters would have had more opportunity to accrete mass at the centre of the potential well, although, it is interesting to note that there are few low luminosity BCGs with small ($<0.01 \,R_{500}$) offsets from the X-ray centroid. 

For comparison with other work, when we analyse the tabulated data of \cite{haarsma2010}, we find no correlation between BCG offset and BCG luminosity for their sample of X-ray luminous clusters. They find that for 90\% of their clusters the BCGs lie within 0.035 of the $R_{500}$ with all cool-core clusters within this distance. However, for our sample this number is only 54\% with 90\% found within 0.10 $R_{500}$. The reason for this difference may relate to the relative flux limits of the samples, with the \cite{haarsma2010} sample containing only massive $M_{\rm cluster}>10^{14}\rm M_{\odot}$, high $L_{X}$ systems. \cite{lin2004} find that 95\% of BCGs in their sample lie within 500\,kpc of the X-ray centroid and we find this number to be 99\% but this is to be expected as our clusters probe down to the group scale where 500\,kpc is comparable to the virial radius.

\begin{figure*}
   \centering
\includegraphics[scale=0.8]{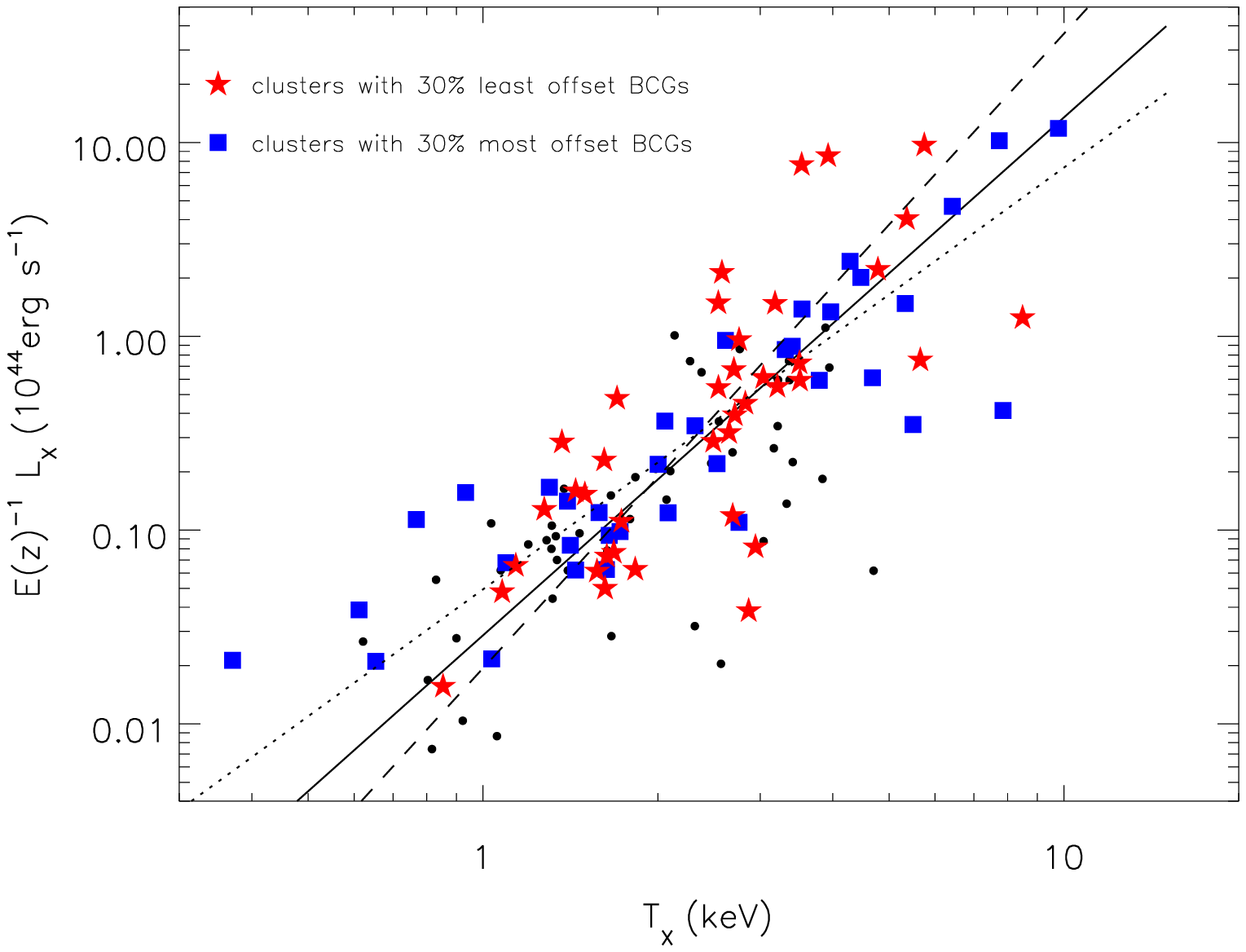} 

\caption[]{The X-ray luminosity plotted against X-ray temperature. Red stars are the clusters containing the 30\% of BCGs least spatially offset from their hosts' X-ray centroids and blue squares are the 30\% most offset. The dashed line is fit to those with the least offset BCGs, the dotted line is fit to those with the most offset BCGs and the solid line is fit to all. The co-located population is found to have a steeper slope than the most offset population. The error bars for the X-ray data can be found in Figure \ref{fig:lxtxerr}.}
   \label{fig:lxtxdist}
\end{figure*}

In Figure \ref{fig:lxtxdist} (the equivalent of Figure \ref{fig:lxtx}) we plot the $L_{X} - T_{X}$ relation for the XCS-SDSS sample. Fits of the form $\log\,(E(z)^{-1}\,L_{X})=a\,\log\,(T_{X}) + b$, with $L_{X}$ in units of $10^{44}\rm erg\,s^{-1}$ and $T_{X}$ in $\rm keV$, are performed to the most offset 30\% of BCGs, with $a_{\rm dist-30}=2.19\pm0.19$ and $b_{\rm dist-30}=-1.30\pm0.19$ and least offset, with $a_{\rm near-30}=3.29\pm0.51$ and $b_{\rm near-30}=-1.72\pm0.18$. 

In a situation mirroring that of the stellar mass of the BCG, the $L_{X} - T_{X}$ relation for the least offset BCGs appears steeper than for the most offset BCGs to a significance of $2.1\sigma$. There is therefore some evidence that above $T_{X}\sim2 \rm \,keV$ the least offset BCGs live in clusters with a higher $L_{X}$ value for a given $T_{X}$ than those that host the most offset BCGs, again the fits to the two populations cross at $T_{X}\sim2\rm \,keV$. Clusters that host the most offset BCGs seem to follow a relation in agreement with the self-similar, $a=2$, case. These results are summarised in Table~\ref{tab:lxtx}.

We speculate that the influence of the dynamical state is connected with strong cool-core clusters as they are also found to have BCGs co-located with their X-ray centroids \citep{sanderson2009,haarsma2010,zhang2011} and have a steeper $L_{X} - T_{X}$ relation than the non-cool-core systems, with $L_{X}\propto T_{X}^{3.33\pm0.15} $ for strong cool-cores and $L_{X}\propto T_{X}^{2.42\pm0.21}$ for non-cool-cores, in agreement with our findings for BCG offset \citep{mittal2011}. The XCS is a serendipitous survey, it is based on cluster images taken over a wide range of off-axis angles (i.e. offsets from the instrument aim point) and so has degraded spatial resolution compared to a  cluster survey based on targeted XMM follow-up (the PSF is strongly off-axis dependent). Therefore, we are unable to determine which of the 123 clusters in our study have cool cores, using spatial fitting. However, it seems a good assumption that a significant fraction of the least offset BCGs will be in cool-core clusters, given the findings of \cite{sanderson2009} and \cite{haarsma2010}. 

A further signature of the dynamical state of the cluster is the luminosity gap between the BCG and the next brightest galaxy in the cluster (e.g. \citealt{smith2010}, see \citealt{harrison2012}, for a detailed study of the luminosity gap in the XCS clusters). This luminosity dominance of the BCG over its neighbours should provide clues as to the merger history of the cluster, as a cluster with more than one `BCG' may have undergone a recent merger, whereas one that is dominant over its satellites may be in a more relaxed system. We perform an analysis of the most dominant and least dominant 30\% of BCGs in the Harrison et al. sample within the context of the $L_{X} - T_{X}$ relation, as above. There is a hint of an effect in that the most dominant BCGs have a steeper relation but it is not statistically significant. With the fit parameters $a$ and $b$ as above, we find for the most dominant 30\% of BCGs $a_{\rm dom-30}=3.72\pm1.30$ and $b_{\rm dom-30}=-1.90\pm0.44$ and for the least dominant this is $a_{\rm non-dom-30}=2.71\pm0.47$ and $b_{\rm non-dom-30}=-1.64\pm0.15$. 

\subsection{Radio-loud BCGs}
\label{sec:rl}

\begin{figure}
   \centering
\includegraphics[scale=0.5]{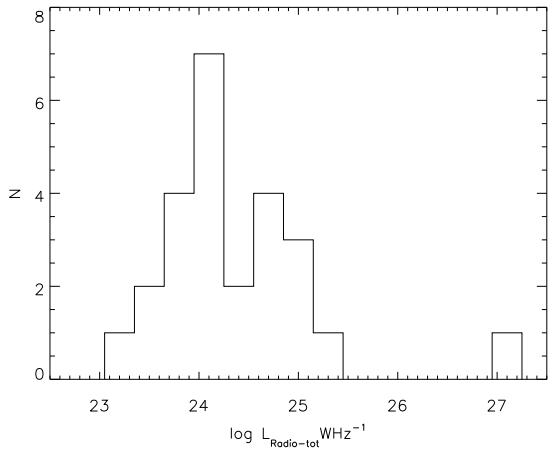} 

\caption{The radio luminosity distribution for the radio-loud sample ($L_{\rm Radio-cen}>2\times10^{23}\rm W\,Hz^{-1}$).}
   \label{fig:radhist}
\end{figure}

As described in \S\ref{sec:data}, we study the radio properties of the BCGs in our sample, considering `radio-loud' BCGs to be those with $L_{\rm Radio-cen}>2\times10^{23}\rm W\,Hz^{-1}$, due to the flux limited nature of the FIRST survey (see Figure \ref{fig:LxLiLrz}, {\it lower panel}). The fraction of radio-loud BCGs in our sample is $0.24\pm0.05$, which is similar to the fraction found in the X-ray selected sample of \cite{lin2007} (0.33) and in the optically selected sample of \cite{best2007} ($0.21\pm0.02$ for clusters with velocity dispersions $\sigma_{v}>500 \rm \,km s^{-1}$). As noted in \S\ref{sec:data} these two similar samples define radio-loud at a fainter limit of $L_{\rm Radio}\gtrsim1\times10^{23}\rm W\,Hz^{-1}$ but in the high stellar mass regime this fraction is unaffected by such a small change in the radio luminosity limit \citep{best2005}, therefore we do not expect this to significantly affect any comparisons. In Figure \ref{fig:radhist} we plot the distribution of radio luminosities of the sample which, due to the flux limit, peaks around $L_{\rm Radio-tot}=10^{24}\rm W\,Hz^{-1}$ with a maximum value of $L_{\rm Radio-tot}=1.2\times10^{27}\rm W\,Hz^{-1}$.

\begin{figure}
   \centering
\includegraphics[scale=0.5]{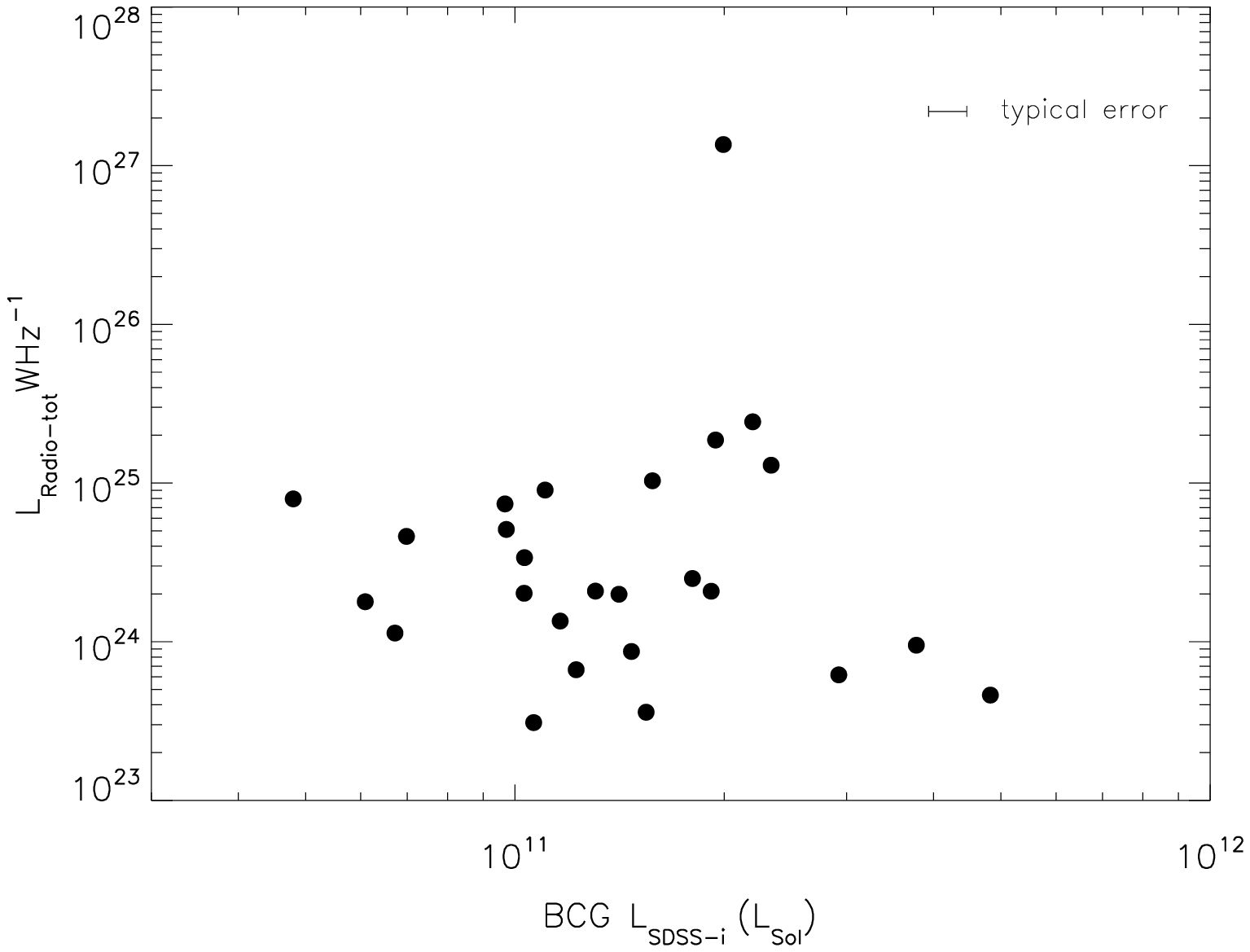} 
\includegraphics[scale=0.5]{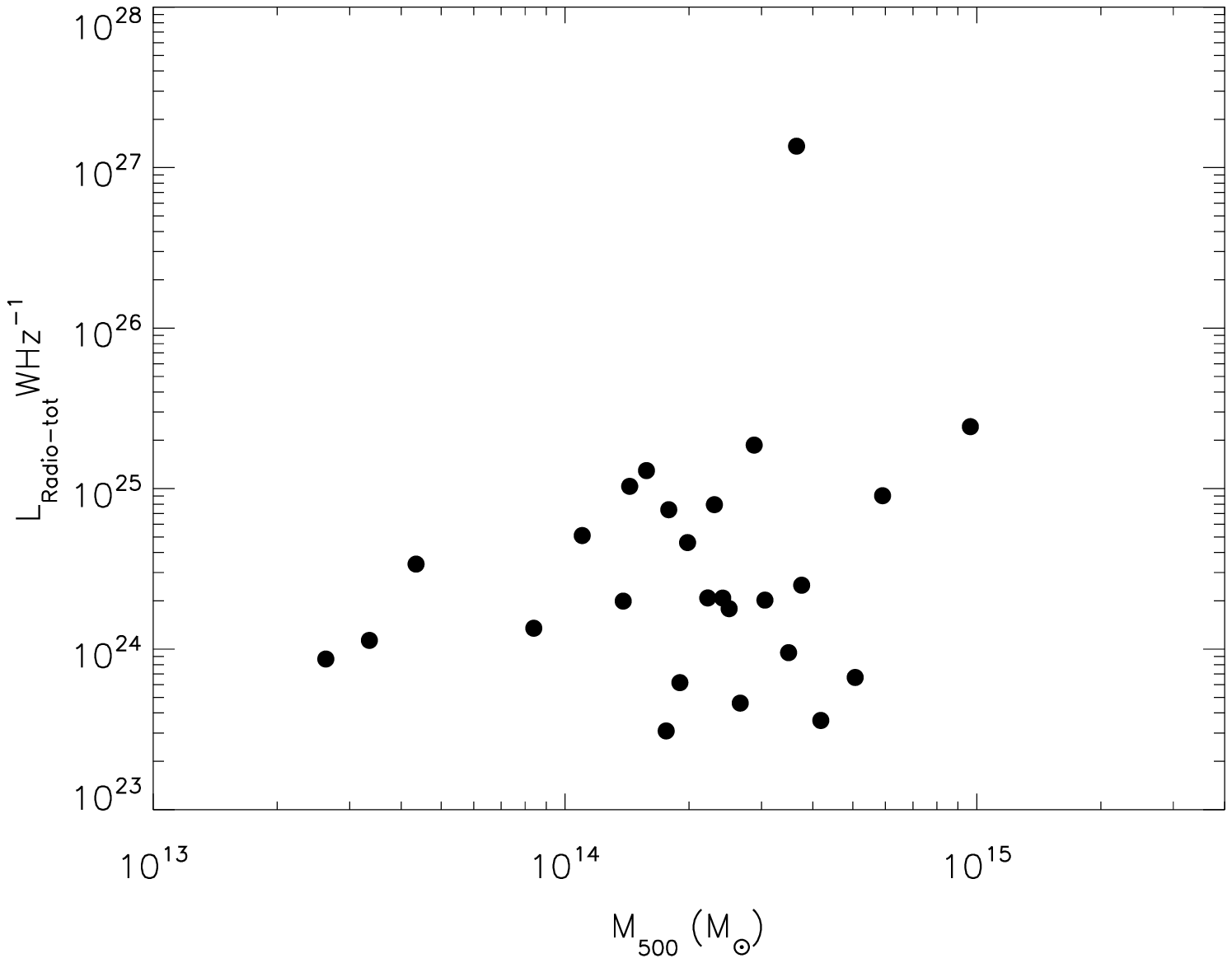} 
\caption{BCG radio luminosity plotted against BCG SDSS $i$-band luminosity (upper) and cluster mass (lower), both showing a lack of significant correlation.}
   \label{fig:lilr}
\end{figure}
\begin{figure}
   \centering
\includegraphics[scale=0.5]{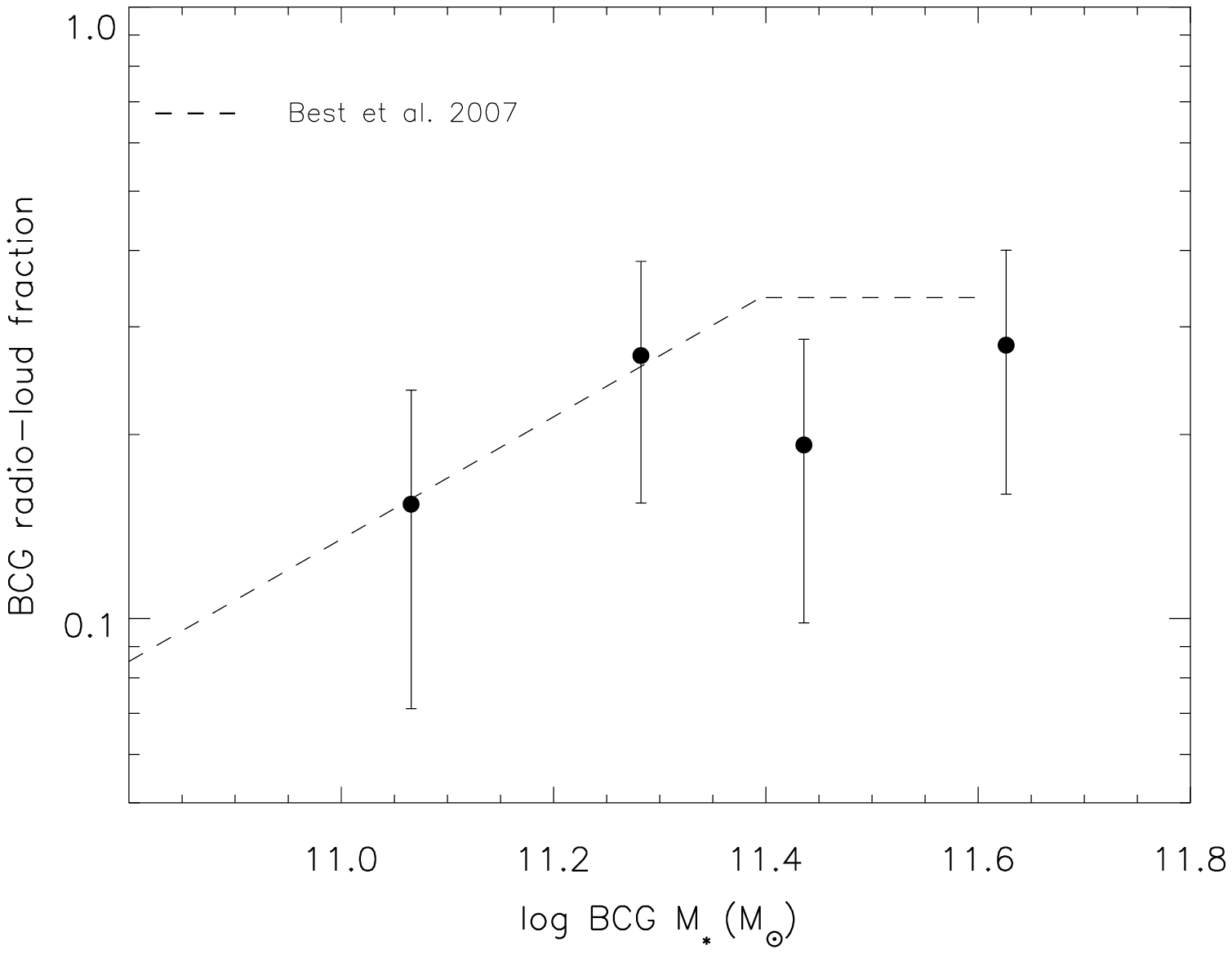} 


\includegraphics[scale=0.5]{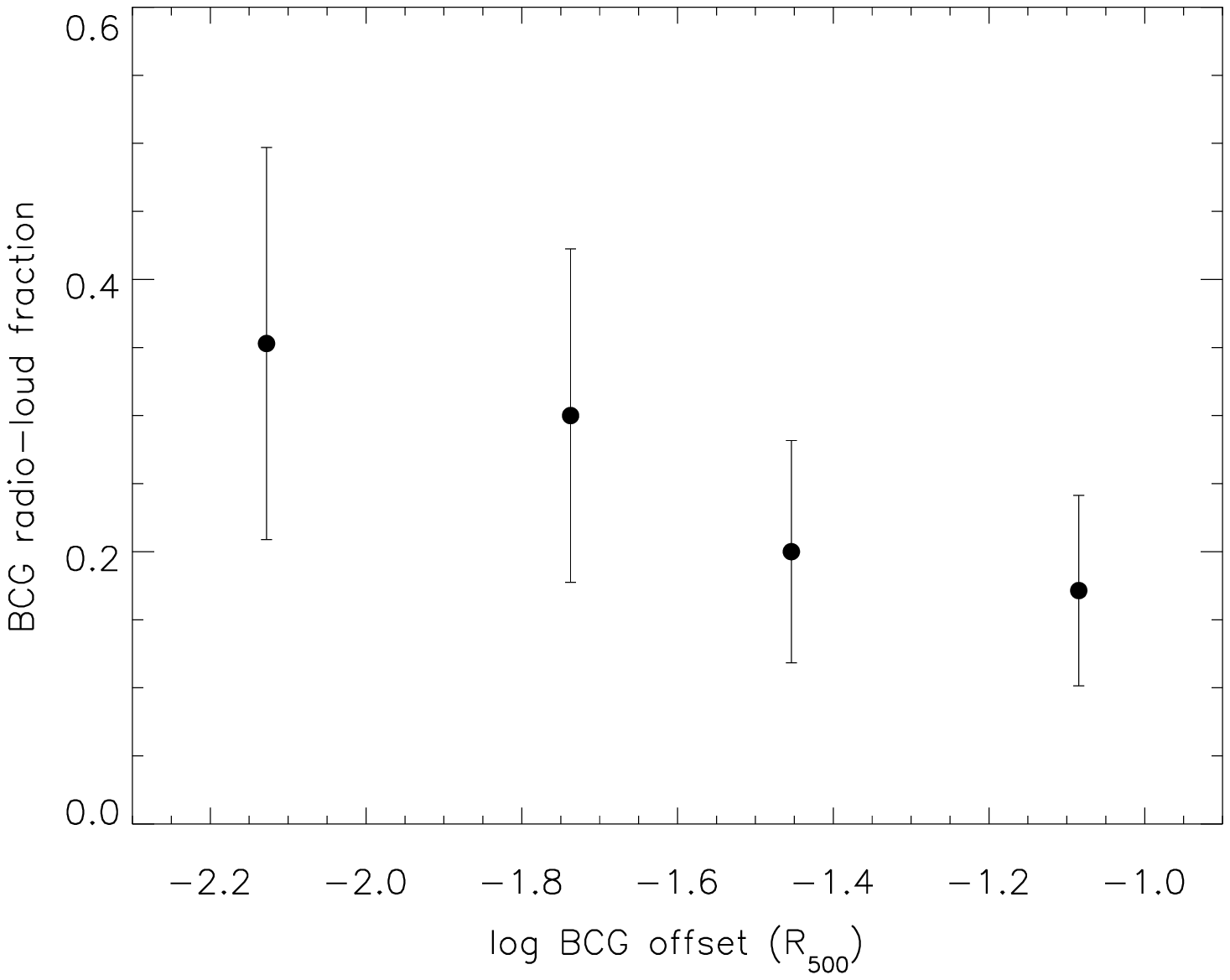} 


\includegraphics[scale=0.5]{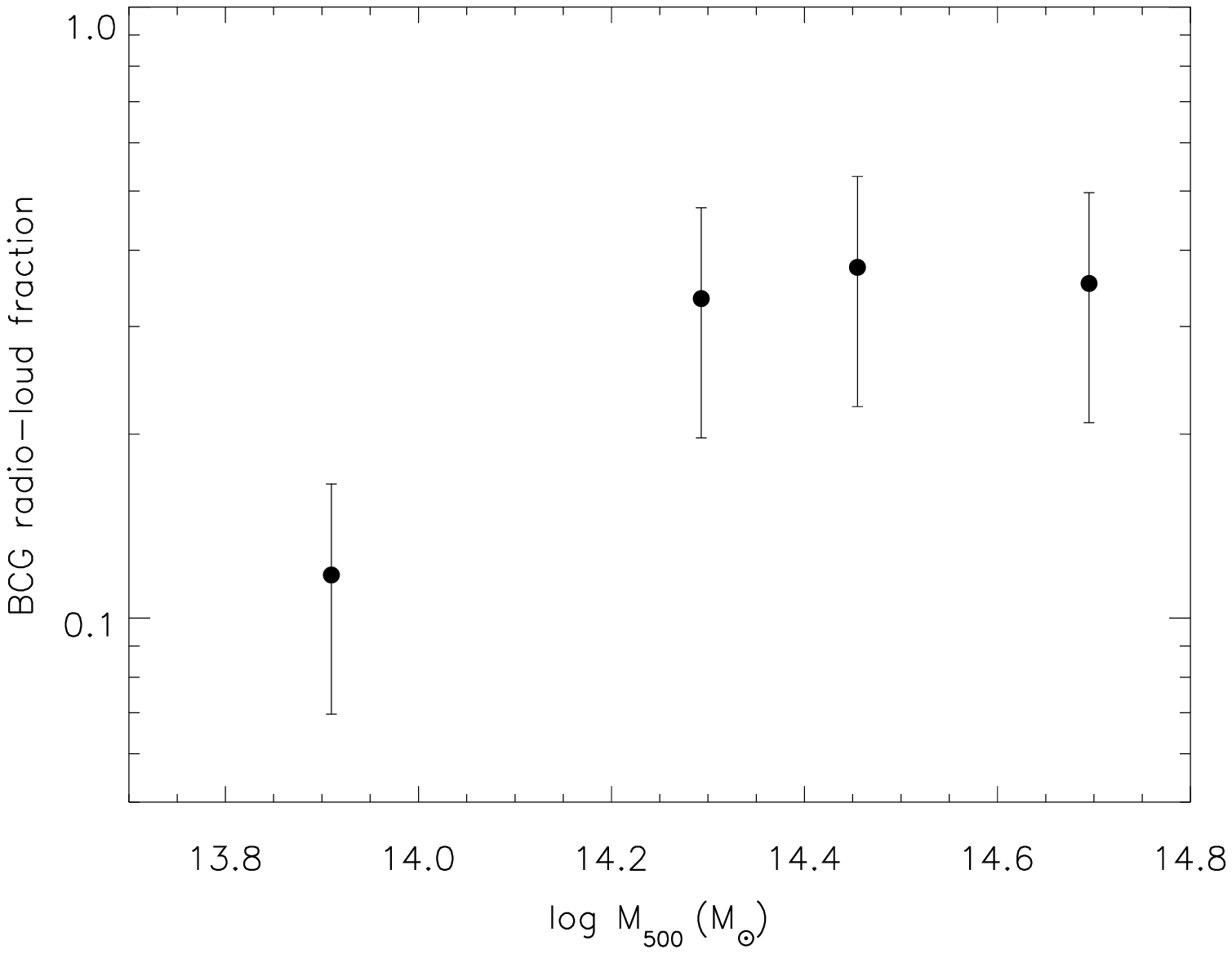} 

   
\caption[]{{\it Upper:} The fraction of radio-loud BCGs per stellar mass bin for the XCS-SDSS-FIRST sample. The dashed line represents the correlation and subsequent flattening from \cite{best2007}, which is in good agreement with our findings. 
{\it Middle:} The fraction of radio-loud BCGs per bin in BCG offset from the X-ray centroid of the cluster for the XCS-SDSS-FIRST sample, which indicates that BCG radio-loud fraction increases when the BCG is co-located with the peak in the ICM surface brightness.
{\it Lower:} The fraction of radio-loud BCGs per cluster mass bin for the XCS-SDSS-FIRST sample. The BCG radio-loud fraction increases significantly with cluster mass.}
\label{fig:rlf}
\end{figure}

There is no evidence for a correlation between the total radio luminosity of the BCG and its optical luminosity or the mass of the cluster (Figure \ref{fig:lilr}). One might expect to find a correlation between BCG $L_{i-\rm SDSS}$ and radio luminosity as stellar mass is thought to correlate with black hole mass via the $M-\sigma$ relation and the accretion rate could thus be higher. In other studies of AGN in BCGs the maximum radio luminosity for a given stellar mass is a function of mass, but the relation for the entire sample has a large scatter, due to the duty-cycle \citep{lin2007}. In Figure \ref{fig:rlf} ({\it upper panel}) we plot the fraction of radio-loud BCGs against BCG stellar mass but find only a weak dependence. We note that \cite{best2007} find the BCG radio-loud fraction  $f_{\rm RL}\propto M_{*}^{1.0}$ when lower BCG masses are included and then a flattening off at high mass. So we can say that for our sample, BCG radio-loud fraction does not correlate with stellar mass but this is still consistent with the measurements of \cite{best2007}, because of the high stellar mass range considered. If we compare the radio-loud fractions above and below the median stellar mass of the sample ($2.3\times10^{11}\rm M_{\odot}$) then the most massive BCGs have $f_{\rm RL}=0.27\pm0.07$ and the least massive BCGs have a radio-loud fraction of $0.21\pm0.06$ so again there is no significant difference, demonstrating that stellar mass has very little influence on radio-loud fraction at these high stellar masses.

There is also no correlation between the total radio luminosity of the BCG and its distance from the X-ray centroid. This is perhaps surprising as one might expect BCGs in relaxed clusters, being located closest to the centre of the cooling gas reservoir, would have more opportunity for fuel to be effectively fed to the black hole, but the duty-cycle may again be responsible for this result. However, in Figure \ref{fig:rlf} ({\it middle panel}) we plot radio-loud fraction against distance from the cluster X-ray centroid and find an excess towards the centre, suggesting a higher fraction of AGN hosting BCGs, in line with the hypothesis above. The fraction of radio-loud BCGs with an offset less than the median offset of 0.03$R_{500}$ $f_{\rm RL}=0.37\pm0.09$ compared to $0.14\pm0.05$ above this, a difference with a significance of $2.3\sigma$ which indicates there is some connection between the two.

\begin{figure*}
   \centering
\includegraphics[scale=0.8]{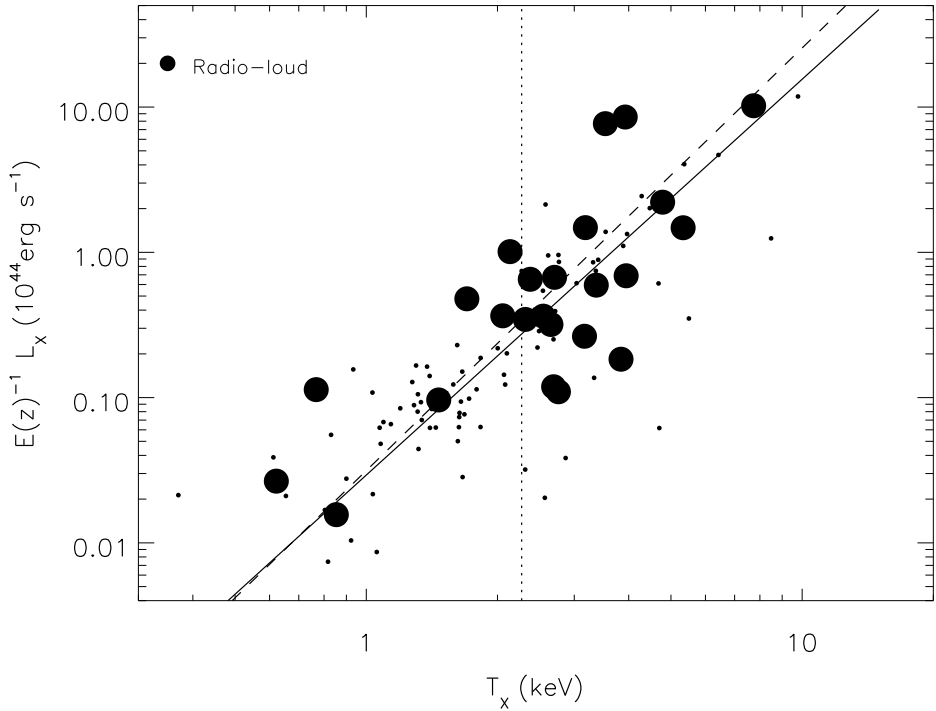} 

\caption{X-ray luminosity plotted against X-ray temperature for the XCS-SDSS-FIRST sample. Large black points are radio-loud ($L_{\rm Radio-cen}>2\times10^{23}\rm W\,Hz^{-1}$) whereas small black dots are below this threshold. The dashed line is a fit to the radio-loud sample and the solid line to the whole sample. The vertical dotted line represent the median $T_{X}$ of the sample. The error bars for the X-ray data can be found in Figure \ref{fig:lxtxerr}.}
   \label{fig:lxtxrad}
\end{figure*}

Figure \ref{fig:lxtxrad} shows the $L_{X} - T_{X}$ relationship for the XCS-SDSS-FIRST sample and we find a fit to this relation of the form $\log\,(E(z)^{-1}\,L_{X})=a\,\log\,(T_{X}) + b$, where $a=2.73\pm0.20$ and $b=-1.53\pm0.08$ in agreement with the fit to the XCS-SDSS sample in Figure \ref{fig:lxtx}. For the radio-loud sample we find that $a_{\rm RL}=2.91\pm0.45$ and $b_{\rm RL}=-1.50\pm0.21$ in agreement with that of the entire sample. These results are summarised in Table \ref{tab:lxtx}. We include a dividing line at $T_{X}=2.1\rm \rm \,keV$ which corresponds to the median $T_{X}$ of the sample.

The fraction of radio-loud BCGs in clusters with $T_{X}<2.1\rm \rm \,keV$ is $f_{\rm RL}=0.10\pm0.04$ and for $T_{X}>2.1\rm \rm \,keV$ it is $f_{\rm RL}=0.38\pm0.09$, a difference of $3\sigma$, demonstrating that cluster mass is an indicator of BCG radio-loudness (see Figure \ref{fig:rlf}, {\it lower panel}). We note that from scaling relations such as that in \cite{xue2000} this value of $T_{X}=2.1\rm \rm \,keV$ (from the $T_{X} - M_{500}$ scaling relation used throughout this paper, $M_{500}=1.4\times10^{14} \rm M_{\odot}$) roughly corresponds to $\sigma_{v}=500\rm \,km s^{-1}$, the value of cluster velocity dispersion used by \cite{best2007} to separate their sample into high and low mass clusters and the $M_{200}=1.6\times10^{14} \rm M_{\odot}$ division used by \cite{lin2007}. The latter also find a significant increase in BCG radio-loud fraction with cluster mass as $f_{\rm RL}=0.36\pm0.03$ in high mass clusters and $f_{\rm RL}=0.13\pm0.05$ in low mass systems, in excellent agreement with our findings. However, \cite{best2007} do not find a significant difference between the radio-loud fractions in their high- and low-mass cluster samples. The most probable reason for this discrepancy is that the \cite{best2007} sample is optically rather than X-ray selected. This may mean that their sample includes a number of clusters which are perhaps under-luminous in X-rays for their optical richness and as such do not make it into our sample.  Alternatively, as their clusters are selected on optical richness, in order to measure a velocity dispersion, they are potentially only selecting the richest of the low mass systems which are perhaps more likely to host AGN. We note that we find no evolution in the AGN fraction with redshift, as for the low redshift half of our sample $z<0.2$ $f_{\rm RL}=0.18\pm0.06$ whereas at $z>0.2$ this number becomes $f_{\rm RL}=0.31\pm0.08$, a difference of only $1.3\sigma$. Table \ref{tab:rlfrac} summarises the radio-loud fraction dependencies.

To crudely transform these radio-loud fractions into estimates of AGN life-times using the method of \cite{lin2007}, we can say that we know there is compelling evidence that BCGs are fully formed by $z=1.5$ \citep{stott2010} making them 6.9\,Gyr old at the median redshift of the survey, $z=0.2$. This may be conservative as the BCG stellar populations formed at $z\gtrsim3$ \citep{stott2010}, although potentially in separate sub-units \citep{delucia2007} and the AGN fraction in clusters, and perhaps therefore duty-cycle, increases with redshift \citep{martini2009}. The low mass, $T_{X}\lesssim2\rm\, keV$ ($M_{500}\lesssim10^{14} \rm M_{\odot}$, $\sigma_{v}\lesssim500\rm \,km s^{-1}$) clusters therefore have life-time estimates of 0.7\,Gyr and the high mass $T_{X}\gtrsim2\, \rm keV$ ($M_{500}\gtrsim10^{14} \rm M_{\odot}$, $\sigma_{v}\gtrsim500\rm km s^{-1}$)  clusters have life-times of 2.7\,Gyr. And for the entire population the life-time is 1.7\,Gyr.

\begin{table*}
\begin{center}
\caption[]{$L_{X}-T_{X}$ relation summary of the form $\log\,L_{X}=a\,\log\,T_{X}+b$, with $L_{X}$ in units of $10^{44} \rm erg\,s^{-1}$, and $T_{X}$ in keV. }
\label{tab:lxtx}
\small\begin{tabular}{lll}
\hline
Sample& $a$  & $b$\\
\hline
XCS-SDSS ALL 				&2.67$\pm$0.19&-1.54$\pm$0.07\\
XCS-SDSS $T_{X}<2.1\,\rm keV$			&2.47$\pm$0.52&-1.39$\pm$0.08\\
XCS-SDSS 30\% most luminous BCGs	&3.69$\pm$0.49&-2.02$\pm$0.24\\
XCS-SDSS 30\% least luminous BCGs	&1.92$\pm$0.33&-1.45$\pm$0.11\\
XCS-SDSS 30\% most offset BCGs		&2.19$\pm$0.19&-1.30$\pm$0.19\\
XCS-SDSS  30\% least offset BCGs		&3.29$\pm$0.51&-1.72$\pm$0.18\\
XCS-HARRISON 30\% most dominant BCGs	&3.72$\pm$1.30&-1.90$\pm$0.44\\
XCS-HARRISON 30\% least dominant BCGs	&2.71$\pm$0.47&-1.64$\pm$0.15\\
XCS-SDSS-FIRST ALL				&2.73$\pm$0.20&-1.53$\pm$0.08\\
XCS-SDSS-FIRST radio-loud			&2.91$\pm$0.45&-1.50$\pm$0.21\\
OWLS ALL					&1.88$\pm$0.16&-1.70$\pm$0.07\\
OWLS 30\% most luminous BCGs	&2.54$\pm$0.21&-1.45$\pm$0.06\\
OWLS 30\% least luminous BCGs	&1.24$\pm$0.13&-1.98$\pm$0.08\\
\hline
\end{tabular}
\end{center}
\end{table*}

\begin{table*}
\begin{center}
\caption[]{BCG radio-loud fractions above and below median(X)}
\label{tab:rlfrac}
\small\begin{tabular}{llll}
\hline
$X$ & median($X$)  & $f_{\rm RL}<$\,median($X$)  & $f_{\rm RL}>$\,median($X$)\\
\hline
BCG stellar mass ($\rm M_{\odot}$)				&$2.3\times10^{11}$&0.21$\pm$0.06&0.27$\pm$0.07\\
BCG offset ($R_{500}$)					&0.03&0.37$\pm$0.09&0.14$\pm$0.05\\
Cluster $T_{X} (\rm keV)$					&2.1&0.10$\pm$0.04&0.38$\pm$0.09\\

\hline
\end{tabular}
\end{center}
\end{table*}

\section{Discussion}
\label{sec:disc}
\subsection{BCG AGN energy input into ICM}
\label{sec:energy}
To study the influence of the AGN on the ICM, we can compare rough estimates of the energy output of the AGN with the thermal energy of the ICM, following a similar prescription to that described in \cite{birzan2004} and \cite{lin2007}. We see no significant correlation between BCG radio luminosity and cluster mass (Figure \ref{fig:lilr}) so, with the caveat of the lower duty-cycle (see \S\ref{sec:rl}), we may expect that AGN outbursts will be more influential on group scales. The mechanical power of the radio jet which causes `bubbles' in the ICM is given by $400\nu\eta L_{\rm Radio-tot}$, where $\nu$ is the frequency (1.4GHz), $\eta$ is the efficiency coefficient and 400 is a factor that comes from the assumption that the gas contained in the outflows is relativistic and adiabatic \citep{birzan2004}. This power can be multiplied by the life-time of the AGN to give the integrated energy input over its history, $E_{\rm AGN}$. For the average radio-loud AGN in our sample with median $L_{\rm Radio-tot}=2.1\times10^{24}\rm W\,Hz^{-1}$ and a life-time of 1.7\,Gyr, $E_{\rm AGN}=6.3\eta\times10^{52}\rm J$.

The thermal energy of the ICM is given by:
\begin{equation}
E_{\rm ICM}=(3/2)k_{\rm B}T_{X}f_{\rm ICM}M_{500}/\mu m_{\rm p}, 
\end{equation}
where $f_{\rm ICM}$ is the ICM mass fraction, the mean molecular weight $\mu=0.59$ and $m_{\rm p}$ is the proton mass. For the median cluster in our sample $M_{500}=1.4\times10^{14}\rm M_{\odot}$ which corresponds to an $f_{\rm ICM}\sim0.09$ of \cite{gonzalez2007} giving $E_{\rm ICM}=8.6\times10^{54}\rm J$.

The ratio of the energy input of the AGN to the thermal energy in the whole system for an average cluster in our sample is therefore $E_{\rm AGN}/E_{\rm ICM}=0.007\eta$ which means that for a $\sim1\times10^{14}\rm M_{\odot}$ cluster the AGN has little effect on the extended ICM. However, this value can be a factor of $\sim2.5$ higher in the centres of clusters (10\% of virial radius) and the value of $\eta$ can also vary from 1/30 - 20 and may change with time \citep{birzan2004,lin2007}. The energy input of the AGN can potentially be a significant fraction of the thermal energy of the ICM at the centre of such a cluster. 

For a typical massive, $M_{500}=1\times10^{15}\rm M_{\odot}$, cluster containing a BCG with the median $L_{\rm Radio-tot}=2.1\times10^{24}\rm W\,Hz^{-1}$ of our sample (as there is no evidence for an $L_{\rm Radio-tot}-M_{500}$ dependency, Figure \ref{fig:lilr}) but with the enhanced life-time of our high cluster mass sample, found in \S\ref{sec:rl} to be 2.7\,Gyr, and with a corresponding $T_{X}\sim8.0\rm \,keV$ and $f_{\rm ICM}=0.13$ \citep{gonzalez2007} we have $E_{\rm AGN}/E_{\rm ICM}=0.0003\eta$. This is a factor of $\sim20$ smaller than the AGN effect on the median mass cluster stated above.

Alternatively, for a typical low mass, $M_{500}=1\times10^{13}\rm M_{\odot}$, group again containing a BCG with the median $L_{\rm Radio-tot}$ of our sample but with the lower life-time of our low cluster mass sample found in \S\ref{sec:rl} to be 0.7\,Gyr and with a corresponding $T_{X}\sim0.3\rm \,keV$ and $f_{\rm ICM}=0.07$ \citep{gonzalez2007} we have $E_{\rm AGN}/E_{\rm ICM}=0.4\eta$. This is comparable to the thermal energy of the entire ICM and thus the AGN can have a major influence on the gas physics of the group. 

\begin{figure}
   \centering
\includegraphics[scale=0.5]{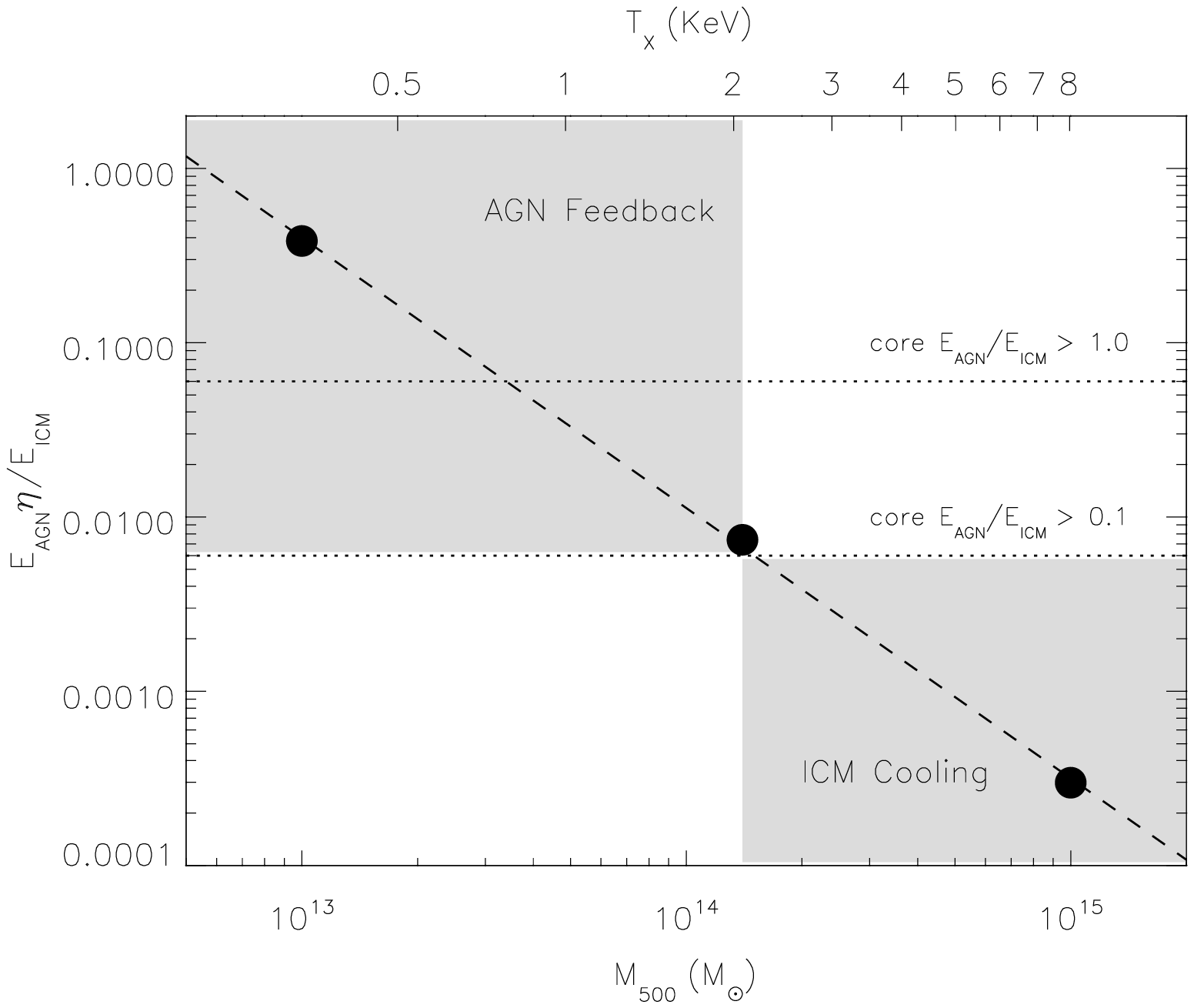} 

\caption[]{This plot is provided as an illustration of the energetics derived from our simplified version of the model taken from \cite{birzan2004} and \cite{lin2007}. The black points are the ratio of integrated AGN energy to the thermal energy of the ICM plotted against cluster mass, assuming the same radio luminosity ($L_{\rm Radio-tot}=2.1\times10^{24}\rm W\,Hz^{-1}$) across all masses. $\eta$ is the efficiency coefficient and the dashed line is a fit to the points. We speculate that when the energy injection from the AGN reaches 10\% of the thermal energy in the core, it can begin to have a significant influence on the ICM and assume this value is reached at $T_{X}\sim 2\,\rm keV$ which corresponds to the crossover between the steepened and self-similar-like relations in \S\ref{sec:lilxtx} and \S\ref{sec:dyn}. We therefore estimate that $\eta\sim5.5$ on average so that, when in combination with a $\sim2.5\times$ heating enhancement in the central region \citep{lin2007}, $E_{\rm AGN}/E_{\rm ICM}>0.1$ at $T_{X}<2\rm\,keV$. Therefore the lower horizontal dotted line represents an estimate of where $E_{\rm AGN}/E_{\rm ICM}>0.1$ in cluster cores. The upper horizontal dotted line represents an estimate of where $E_{\rm AGN}/E_{\rm ICM}>1.0$ in cluster cores, which may lead to a significant disruption of the ICM at $T_{X}<1\,\rm keV$. The shaded regions represent where ICM cooling dominates and where AGN feedback becomes influential, for this simple treatment.}
   \label{fig:eratio}
\end{figure}

The results of this over-simplified analysis are plotted for illustration in Figure \ref{fig:eratio}. From this plot we can see that a central radio source with the same radio luminosity in all clusters has significantly more effect on group scales than in massive clusters, with the AGN able to inject a significant fraction of the thermal energy of the ICM in the lowest mass systems. A similar trend is also found by \cite{ma2011} when studying the energetics of AGN over a range of redshifts. 

If we consider the transition from AGN feedback to cooling dominated clusters in the $L_{X}-T_{X}$ relations, discussed in \S\ref{sec:lilxtx} and \S\ref{sec:dyn}, that appears to take place at $T_{X}\sim2\rm\,keV$ (see \S\ref{sec:bcgopt}) in the context of Figure \ref{fig:eratio}, then an average value of $\eta=5.5$ takes $E_{\rm AGN}/E_{\rm ICM}$ to $\gtrsim0.04$ at the corresponding mass $\sim1-2\times10^{14}\rm M_{\odot}$. If the value is boosted by a factor of $\sim2.5$ in cluster cores \citep{lin2007} then $E_{\rm AGN}/E_{\rm ICM}>0.1$ at masses below this and AGN feedback can start to contribute significantly to the core energetics, with $E_{\rm AGN}/E_{\rm ICM}>1.0$ in the cores of $M_{500}<4\times10^{13}\rm M_{\odot}$ ($T_{X}<0.8\,\rm keV$) systems which could remove the central gas entirely. This simple energetic argument may explain the transition from cooling to AGN feedback dominance seen in \S\ref{sec:lilxtx} and \S\ref{sec:dyn} and discussed in \S\ref{sec:bcgopt}. 

\subsection{Time averaged effect of AGN}
\label{sec:bcgopt}
The scaling relations between the BCG stellar properties and the X-ray properties of the ICM demonstrate that central galaxy mass correlates with the mass of its host dark matter halo over the range $10^{13}<M_{500}<10^{15} \rm M_{\odot}$. The steep relation we find is similar to that found by \cite{mittal2009} who use an almost identical definition of magnitude and the same fitting method. The agreement between our observed BCG - cluster scaling relations and those from the OWLS simulation with AGN feedback \citep{mccarthy2010} may mean that AGN feedback, invoked to regulate gas on both the galaxy and cluster scales, is indeed the dominant baryonic feedback process at work in dense environments. We note that the OWLS simulations indicate that ejection of low entropy gas at high redshift from $L^{\star}$ galaxies is key to producing the dramatic change in the $L_{X}-T_{X}$ relation seen in Figure \ref{fig:owls}, with the cooling of the gas at late times regulated by feedback from the BCG AGN.

The lack of a significant correlation between BCG stellar mass and its distance from the X-ray centroid adds weight to the observation that BCGs form early $z\gg1$ \citep{collins2009,stott2010,stott2011} and that subsequent disturbances to the cluster, perhaps via cluster-cluster merging, thus have little influence on BCG mass.

The interrelation between the BCG properties and the ICM is highlighted in the fits to the $L_{X} - T_{X}$ relation for different sub-samples. Fits to clusters which host the most and least massive BCGs are discrepant to a significance of $3\sigma$. The most massive BCGs live in clusters with a steepened relation, such that above $T_{X}\sim2\rm \,keV$ ($M_{500}\sim1\times10^{14} \rm M_{\odot}$) they reside in clusters with relatively high $L_{X}$ compared to those with the lowest mass BCGs and below this they occupy the same region of the $L_{X} - T_{X}$ plot. This is also seen in the OWLS simulations, albeit over a different cluster mass range, further demonstrating that simulations that model AGN feedback are broadly able to recreate observable properties. The offset of the BCG from the X-ray centroid of the cluster has a similar effect on the $L_{X}-T_{X}$ relation, with the BCGs co-located with their clusters' X-ray centroid seemingly having a steeper slope than their more offset counterparts. 

From the literature, other populations that have steepened $L_{X}-T_{X}$ relation slopes are those with extended AGN radio emission and those with a strong cool-core \citep{mag2007, mittal2011}. In a time averaged sense we expect the most massive BCGs in our sample to harbour the most powerful AGN, even if they are not currently accreting. This is because BCG mass correlates with black hole mass through the $M_{\rm BH} - M_{\rm Bulge}$ relation and black hole mass governs the accretion rate which is coupled to the AGN energy output (e.g. \citealt{booth2010}). Having a BCG co-located with the X-ray centroid is also associated with AGN activity and the presence of a cooling core \citep{sanderson2009,haarsma2010,mittal2011,zhang2011}, confirmed for this sample by the decrease in BCG radio-loud fraction with distance from X-ray peak (Figure \ref{fig:rlf}, {\it middle panel}).

As $L_{X}$ is a proxy for the density of the cooling gas, this steepening of the $L_{X}-T_{X}$ relation means that in the cluster mass regime ($T_{X}>2\,\rm keV$) the most massive BCGs, or those co-located with the cluster X-ray centroid, live in systems where gas cooling dominates, whereas in the group mass regime they live in systems where AGN feedback dominates. To explain this, we note that we showed in \S\ref{sec:bcgscale}, that BCG mass depends on cluster mass through a sub-unity exponent and that radio luminosity does not correlate with cluster mass (\S\ref{sec:rl}). Therefore, the effect of AGN feedback will not be self-similar. In \S\ref{sec:energy} we see this in action as, assuming the same radio luminosity of AGN at each cluster mass, we find that its energy becomes increasingly significant with decreasing cluster mass, despite the lower duty-cycle, simply because the gravitational potential well is shallower and the gas mass is lower in groups, making it easier to disturb the ICM and raise its temperature. BCGs co-located with the cluster X-ray centroid, and therefore with the peak in ICM density, will have a more effective supply of gas to feed their central black hole and as such the energetics argument is the same as that for the most massive BCGs.


The clusters with the lowest mass BCGs, or those most offset from their X-ray centroid, have $L_{X}-T_{X}$ relations that are consistent with the self-similar situation of $L_{X}\propto T_{X}^{2}$. So in clusters where there is little evidence that cool-cores and non-gravitational effects such as AGN feedback have an influence on the ICM, the X-ray observables show that this is indeed the case. 

The apparent lack of a departure from self-similarity in the low BCG mass/large BCG offset systems may give insight into their formation or recent merger history. Recent cluster-cluster mergers may have disrupted the potential, removing the opportunity for a cool-core to form, so the BCGs will look under-luminous as they originate from one of the merging clusters and thus do not follow the average BCG mass-cluster mass trend from \S\ref{sec:bcgscale}. A recent disturbance would also explain the offset of the BCG from the X-ray gas centroid. However, we find no correlation between BCG mass and BCG offset in \S\ref{sec:dyn}. Also, when analysing the effect of the luminosity gap of the BCG down to the next brightest members, there is no evidence that clusters with several `BCGs' follow different $L_{X}-T_{X}$ relations. Alternatively, ignoring cluster-cluster merging, perhaps these are systems where a strong mechanical or heating mechanism removed the low entropy gas at an early epoch (e.g. AGN feedback from $L^{\star}$ galaxies at $z\gtrsim1$, \citealt{mccarthy2011}) and this has been regulated ever since by AGN feedback from the BCG, making both the ICM and BCG under-luminous, in the massive ($T_{X}>2\rm \,keV$) clusters, due to the lack of cooling gas available for star formation.   

We note here that a fit to the group $L_{X} - T_{X}$ relation ($T_{X}<2.1\,\rm keV$), gives a slope consistent with that of the whole population ($L_{X}\propto T_{X}^{2.47\pm0.52}$) in contrast with the steepened slope found by \cite{helsdon2000} ($L_{X} \propto T_{X}^{4.9}$). We speculate that this difference is because that sample is optically selected, unlike the X-ray selected sample presented here. This lack of slope change on the group scale can be reconciled with the above results as, even though the gravitational potential is easier to overcome in groups, strong AGN feedback still only occurs in systems where it is required to stop gas over-cooling.

A caveat is that our results are likely affected by selection biases, although our sample has an $L_{X}$ limit reasonably constant with $z$. If a lower portion of the group $L_{X}$ range is missing at low $T_{X}$ below this limit, then we may expect a shallower slope in this regime (i.e. see Figure 5 of \citealt{allen2011}), which would particularly affect the $L_{X}-T_{X}$ relation of those with low mass BCGs. We attempt to test this with the volume limited OWLS simulation as this is qualitatively similar to the observed data. By cutting the data at a $L_{X}=3\times10^{41} \rm erg\,s^{-1}$ to remove the 30\% lowest $L_{X}$ groups and refitting the $L_{X}-T_{X}$ relation for the whole sample and the luminous and faint BCG sub-samples. As one might expect the slopes all become shallower but are still consistent with the values from \S\ref{sec:owls} within $1\sigma$ except for the slope of the faint BCG sample which changes by $\sim1.5\sigma$ and thus does not affect the conclusions. From this we conclude that the faint BCG sample will be affected most by selection bias, although it is difficult to estimate the magnitude of the effect, particularly with the complicated selection function due to the serendipitous nature of the XCS.

\subsection{Instantaneous effect of AGN}
\label{sec:bcgrad}
 
The $L_{X} - T_{X}$ relation fit to the radio-loud AGN is in agreement with that of the whole sample, although we do not have the statistics to go further and say whether this is associated with any of the optically selected sub-samples described in \S\ref{sec:bcgopt}. We find the result that AGN radio luminosity does not correlate with either the stellar mass of the BCG, its offset from the X-ray centroid, or cluster mass. As discussed in the results section, this is most likely due to the AGN duty-cycle which means that AGN are switched on and accreting intermittently and therefore any correlations of this nature are washed out. To understand how AGN relate to their host galaxy and cluster, the AGN fraction is a more useful measure \citep{best2007}.

We find that the AGN fraction for the BCGs in our sample is only weakly dependent on BCG mass for the limited range of high galaxy masses we observe. One might expect to see a correlation as black hole accretion rate is thought to be correlated with the mass of the black hole \citep{bondi1952} and thus the mass of the host galaxy through the $M_{\rm BH} - M_{\rm Bulge}$ relation \citep{magorrian1998}. \cite{best2007} do find a correlation with stellar mass but this is for a sample that encompasses BCGs down to a mass of $4\times10^{10}\rm M_{\odot}$ whereas only 8\% of our sample have stellar mass less than $10^{11}\rm M_{\odot}$. The lack of correlation may indicate that in the cluster environment where all BCGs are massive, the mass of the black hole is not what is driving the AGN activity.

The AGN fraction correlates most strongly with the cluster $T_{X}$ and therefore cluster mass (and similarly with $L_{X}$). The AGN fraction in clusters below the median halo mass ($T_{X}\sim2\,\rm keV$, $\sim10^{14}\rm M_{\odot}$) being lower by $3\sigma$ than those in clusters with a halo mass greater than this. An enhancement in radio-loud fraction is also found for BCGs within the projected median distance from the X-ray centroid compared to those outside this radius ($0.03R_{500}$). Both of these phenomena relate to the amount of cooling gas available as $L_{X}$ correlates with gas density so as in \S\ref{sec:bcgopt}, this suggests that the black hole needs an environment with significant amount of cooling gas available to fuel it. The fact that the small number of radio-loud BCGs in the lowest mass ($T_{X}<1\,\rm keV$) clusters are found in those with average to relatively high $L_{X}$ may back this claim (see Figure \ref{fig:lxtxrad}). As radio-loudness is an indicator of ongoing AGN feedback we may expect these clusters to perhaps lower in $L_{X}$ or increase in $T_{X}$ over the duty-cycle of the feedback process.

From the above relations between AGN fraction and both galaxy and cluster properties it seems that the key parameters that govern the presence of AGN in BCGs are primarily the cluster mass/$T_{X}$/$L_{X}$ and to a lesser extent the BCG offset from the cluster X-ray centroid, but not BCG mass. 
A picture is therefore emerging that the supermassive black holes at the centres of BCGs in cluster cores know more about their host cluster than they do about their host galaxy. While this is consistent with some simulations \citep{booth2010, gaspari2011} and results from the clustering of radio-loud, luminous red galaxies \citep{wake08radio}, it is perhaps puzzling as it means that the sub pc scale is governed by the 100s kpc scale and not the kpc scale. We conclude that the connection must be due to the amount of fuel available and thus what drives AGN activity in the central galaxies of clusters is the availability of a large, cooling gas supply that can be effectively fed to the central black hole. This explains the cool-core AGN connection and the break from self-similarity for these sub-populations. How this cooling gas on kpc scales is funnelled down into the accretion disk of the black hole remains an open question. 


The correlations found in this paper imply that we can use complementary information about the component galaxies to characterise sub-samples in current and future cluster surveys such as the XCS which compromise on X-ray integration time in favour of large area coverage, in order to improve constraints on cosmological parameters.

\section{Summary}

The XCS first data release \citep{mehrtens2011} is key to this study as it provides an ideal sample of groups and clusters that span a large range in both $L_{X}$ and $T_{X}$, which is essential to probe the effects of AGN feedback on the ICM in a large, statistically significant, sample. The results lead us to conclude that $T_{X}\sim2 \rm \,keV$ ($M_{500}\sim10^{14}\rm M_{\odot}$, $L_{X}\sim10^{43} \rm erg\,s^{-1}$, $\sigma_{v}\sim500\rm km s^{-1}$) is an important mass scale which delineates the boundary between the cooling dominated systems above this mass and the AGN feedback dominated systems below. We propose that this is an excellent empirical demarcation between what is referred to as a `group' and what is a `cluster' in agreement with previous, perhaps ad hoc, definitions (e.g. \citealt{bower2004}) and similar to the break in the entropy-at-a-radius-enclosing-a-given-mass vs. temperature relations of \cite{ponman2003}. Recent results indicate that AGN cavity power exceeds the luminosity of the cooling ICM at a similar mass scale \citep{osullivan2011,gitti2011}. 

We summarise our key findings as follows:

\begin{itemize}

\item[1.] The $L_{X} - T_{X}$ relation for the clusters with the most massive BCGs is significantly steeper than for those with the least massive BCGs, with a similar story for those least and most offset from the peak in the ICM emission. This must relate to both AGN and cool-core phenomena as similarly steep relations are found for these sub-samples in the literature.

\item[2.] Clusters without a massive, centrally located BCG instead follow self-similarity, confirming that it is the complex interplay between cooling gas and the central AGN that leads to the departure from simple gravitational heating. We speculate that the self-similar clusters may have a different feedback, formation or recent merger history which stops gas from cooling onto the BCG's black hole.

\item[3.] The steepened and self-similar $L_{X} - T_{X}$ relations described in points 1. and 2. cross at $T_{X}=2\rm \,keV$ and thus we conclude that gas cooling dominates above this mass scale and AGN feedback dominates below.

\item[4.] We derive BCG-cluster scaling relations over the mass range $10^{13}<M_{500}<10^{15}\rm M_{\odot}$. The $M_{\rm *BCG} - M_{500}$ relation has an exponent $<1$, implying that BCGs do not grow at the same rate as their host dark matter halos. The X-ray scaling relations can be successfully reproduced by cosmological hydrodynamical simulations that include AGN feedback, such as OWLS.

\item[5.] A quarter of the BCGs in our sample are radio-loud, in agreement with similar studies, but we find a strong dependence on environment, in that those in clusters with $T_{X}>2\rm \,keV$ are 4 times more likely to host a radio-loud AGN than those with $T_{X}<2\rm \,keV$. The radio-loud fraction also increases for BCGs closest to the peak in the ICM emission but the stellar mass of the BCG has little effect. This indicates that it is the availability of a fuel supply of cooling gas from the halo environment that governs AGN activity in the most massive galaxies.

\item[6.] From a crude calculation of the energetics to explain points 1., 2. and 3., it is easier for AGN to affect the group scale systems than the cluster scale, which may be as a direct result of point 4., the sub-unity exponent in the relationship between BCG mass (and therefore black hole mass) and the mass of its host cluster combined with the lack of correlation between BCG radio luminosity and cluster mass. 

\item[7.] The XCS sample shows no evidence for the group scale $L_{X} - T_{X}$ slope being different from that of the cluster scale. This can be reconciled with the above results as, even though the gravitational potential is easier to overcome in groups, strong AGN feedback still only occurs in systems where it is required to stop gas over-cooling.

\end{itemize}


\section*{Acknowledgments}

We first thank the anonymous referee for their thorough report, which has improved the clarity of this paper. We acknowledge Jim Geach, Stephen Fine, Sean McGee, James Mullaney and Rich Bielby for useful discussions. J.P.S. acknowledges funding from the Science and Technology Facilities Council. A.K.R., E.J.L.-D., N. M.\ and A.R.L.\ were supported by the Science and Technology Facilities Council [grant numbers ST/F002858/1 and ST/I000976/1]. M.S. acknowledges financial support from the Swedish Research Council (VR) through the Oskar Klein Centre. P.T.P.V. acknowledges financial support from project PTDC/CTE-AST/64711/2006, funded by Funda\c{c}\~{a}o para a Ci\^{e}ncia e a Tecnologia.

Funding for SDSS-III has been provided by the Alfred P. Sloan Foundation, the Participating Institutions, the National Science Foundation, and the U.S. Department of Energy. The SDSS-III web site is http://www.sdss3.org/. SDSS-III is managed by the Astrophysical Research Consortium for the Participating Institutions of the SDSS-III Collaboration including the University of Arizona, the Brazilian Participation Group, Brookhaven National Laboratory, University of Cambridge, University of Florida, the French Participation Group, the German Participation Group, the Instituto de Astrofisica de Canarias, the Michigan State/Notre Dame/JINA Participation Group, Johns Hopkins University, Lawrence Berkeley National Laboratory, Max Planck Institute for Astrophysics, New Mexico State University, New York University, Ohio State University, Pennsylvania State University, University of Portsmouth, Princeton University, the Spanish Participation Group, University of Tokyo, University of Utah, Vanderbilt University, University of Virginia, University of Washington, and Yale University.

\bibliographystyle{mn2e}
\bibliography{agnfb}
\appendix\clearpage
\section{Additional figures}
\begin{figure*}
   \centering
\includegraphics[scale=0.8]{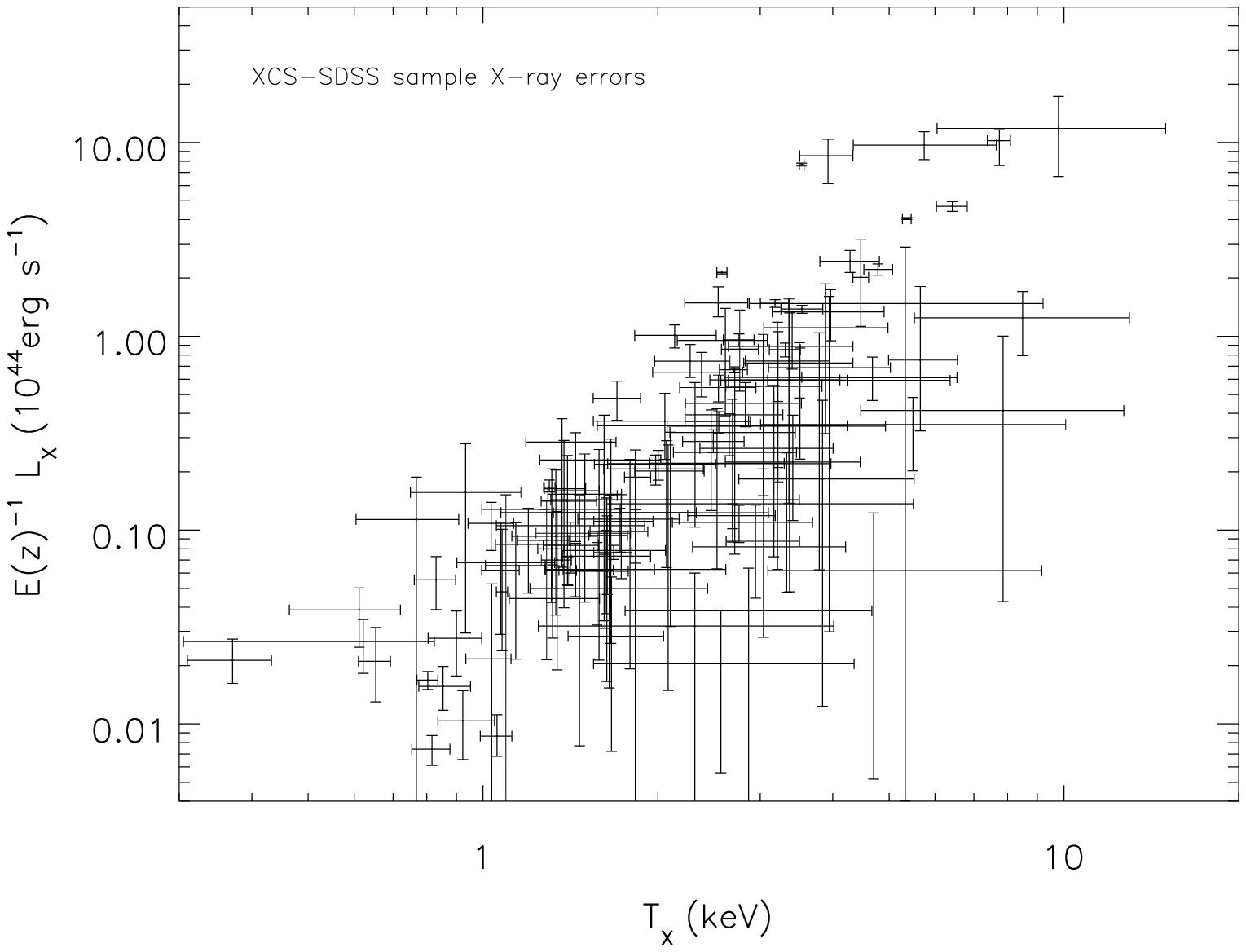} 
\caption{The X-ray luminosity plotted against X-ray temperature for the XCS-SDSS sample to illustrate their associated errors. }
   \label{fig:lxtxerr}
\end{figure*}



\label{lastpage}

\end{document}